\documentclass[aps,superscriptaddress]{revtex4}
\usepackage{amssymb,amsmath,epsfig}
\usepackage[colorlinks=true, pdfstartview=FitV, linkcolor=blue, citecolor=red, urlcolor=magenta, breaklinks=true]{hyperref}
\usepackage{graphicx}
\usepackage{epstopdf}
\usepackage[capposition=top]{floatrow}
\usepackage{subfig}

\begin{document}
\title{Confinement of Fermions in Tachyon Matter at Finite Temperature}

%twocolumn[  \begin{@twocolumnfalse}
\author{Adamu Issifu}\email{ai@academico.ufpb.br}
\affiliation{Departamento de F\'isica, Universidade Federal da Para\'iba, 
Caixa Postal 5008, 58051-970 Jo\~ao Pessoa, Para\'iba, Brazil}

\author{Julio C.M. Rocha}\email{julhocerza@gmail.com}
\affiliation{Departamento de F\'isica, Universidade Federal da Para\'iba, 
Caixa Postal 5008, 58051-970 Jo\~ao Pessoa, Para\'iba, Brazil}

\author{Francisco A. Brito}\email{fabrito@df.ufcg.edu.br}
\affiliation{Departamento de F\'isica, Universidade Federal da Para\'iba, 
Caixa Postal 5008, 58051-970 Jo\~ao Pessoa, Para\'iba, Brazil}
\affiliation{Departamento de F\'{\i}sica, Universidade Federal de Campina Grande
Caixa Postal 10071, 58429-900 Campina Grande, Para\'{\i}ba, Brazil}

\begin{abstract}
We study a phenomenological model that mimics the characteristics of QCD theory at finite temperature. The model involves fermions coupled with a modified Abelian gauge field in a tachyon matter. It reproduces some important QCD features such as, confinement, deconfinement, chiral symmetry and quark-gluon-plasma (QGP) phase transitions. The study may shed light on both light and heavy quark potentials and their string tensions. Flux-tube and Cornell potentials are developed depending on the regime under consideration. Other confining properties such as scalar glueball mass, gluon mass, glueball-meson mixing states, gluon and chiral condensates are exploited as well. The study is focused on two possible regimes, the ultraviolet (UV) and the infrared (IR) regimes. 
\end{abstract}

\maketitle
\pretolerance10000
\section{Introduction}

Confinement of heavy quark states $\bar{Q}Q$ is an important subject in both theoretical and experimental study of high temperature QCD matter and quark-gluon-plasma phase (QGP) \cite{Sarkar}. The production of heavy quarkonia such as the fundamental state of $\bar{c}c$ in the Relativistic Heavy Iron Collider (RHIC) \cite{Adare} and the Large Hadron Collider (LHC) \cite{Aad} provides basics for the study of QGP. Lattice QCD simulations of quarkonium at finite temperature indicates that $J/\psi$ may persists even at $T=1.5T_c$ \cite{Asakawa} i.e. a temperature beyond the deconfinement temperature. However, in simple models such as the one under consideration, confinement is obtained at $T\,<\,T_c$, deconfinement and QGP phase at $T\,\geq\,T_c$. Several approaches of finite temperature QCD has been studied over some period now, but the subject still remain opened because there is no proper understanding on how to consolidate the various different approaches \cite{Umeda,Datta,Jako,Aarts,Ding,Ohno}. Ever since the Debye screening of heavy $\bar{Q}Q$ potential was put forward \cite{Matsui}, leading to the suppression of quarkonia states such as $J/\psi$. Attention has been directed to investigations to understand the behavior of $\bar{Q}Q$ at the deconfined phase by non relativistic calculations \cite{Digal,Shuryak,Cabrera,Alberico,Mocsy} of the effective potential or lattice QCD calculations \cite{Asakawa,Umeda,Datta,Jako,Aarts,Iida}. 

We will study the behavior of light quarks such as up (u), down (d) and strange (s) quarks and also heavy quarks such as charm (c), and other heavier ones with temperature ($T$). We will elaborate on the net confining potential, vector potential, scalar potential energy, string tension, scalar glueball masses, glueball-meson mixing states and gluon condensate with temperature. {Additionally, confinement of quarks at a finite temperature is an important phenomenon for understanding QGP phase. This phase of matter is believed to have formed few milliseconds after the Big Bang before binding to form protons and neutrons. Consequently, the study of this phase of matter is necessary in understanding the early universe. However, creating it in laboratory poses a great challenge to physicists. Because immensely high energy is needed to break the bond between hadrons to form free particles as it existed at the time. Also, the plasma, when formed, has a short lifetime, so they decay quickly to elude detection and analyses. Some theoretical work has been done in tracing the signature of plasma formation in series of heavy ion collisions. Nonetheless, a phenomenological model that throws light on the possible ways of creating this matter is necessary. In this model, the QGP is expected to be created at extremely high energy and density by regulating temperature, $T\,>\,T_c$. Aside, these specific motivations for this paper, all others pointed out in \cite{Issifu,Issifu1} are also relevant to this continuation.} This paper is meant to compliment the first two papers in this series \cite{Issifu,Issifu1}. Particularly, it is a continuation of \cite{Issifu}. We will try as much as possible to maintain the notations in \cite{Issifu} in order to make it easy to connect the two. That notwithstanding, we will redefine some of the terms for better clarity to this paper when necessary. Generally, we will investigate how thermal fluctuations affect confinement of the fermions at various temperature distribution regimes. The Lagrangian adopted here has the same structure as the one used in \cite{Issifu}, so we will not repeat the discussions.

{Even though strong interactions are known to be born out of non-Abelian gauge field theory, no one has been able to compute confining potentials through this formalism. Therefore, several phenomenological models have been relied upon to describe this phenomenon. Common among which are linear and Cornell potential models. Bound states of heavy quarks form part of the most relevant parameters for understanding QCD in high-energy hadronic collisions together with the properties of QGP. As a results, several experimental groups such as CLEO, LEP, CDF, D0, Belle and NA5 have produced data and currently ongoing experiments at BaBar, ATLAS, CMS and LHCb are producing and are expected to produce more precise data in the near future \cite{Soni,Eichten,Godfrey,Barnes,Brambilla,Brambilla1,Andronic}.}

The paper is organized as follows: In Sec.~\ref{TM} we present the model and further calculate the associated potentials and string tensions in the ultraviolet (UV) and the infrared (IR) regimes. In Sec.~\ref{TF} we present the technique for introducing temperature into the model. We divide section ~\ref{VSG} into three subsections where we elaborate on vector potential in Sec.~\ref{VP}, scalar potential energy in Sec.~\ref{SP}, the quark and gluon condensates in Sec.~\ref{QGC} and chiral condensate in Sec.~\ref{CC}. In Sec.~\ref{CON} we present our findings and conclusions.   

\section{The model}\label{TM}
We start with the Lagrangian density 
\begin{equation}\label{9}
\mathcal{L}=-\dfrac{1}{4}G(\phi)F_{\mu\nu}F^{\mu\nu}+\dfrac{1}{2}\partial_{\mu}\phi\partial^{\mu}\phi-V(\phi)-\bar{\psi}\left(i \gamma^{\mu}\partial_{\mu}+q \gamma^{\mu}A_{\mu}-m_{q}G(\phi)\right)\psi,
\end{equation}
with a potential given as 
\begin{equation}\label{a1}
V(\phi)=\frac{1}{2}[(\alpha\phi)^2-1]^2.
\end{equation}
The Lagrangian together with the potential and tachyon condensation governed by the color dielectric function $G(\phi)$ and dynamics of the scalar field produces the required QCD characteristics.

The equations of motion are
\begin{equation}\label{11}
 \partial_{\mu}\partial^{\mu}\phi+\dfrac{1}{4}\dfrac{\partial G(\phi)}{\partial\phi}F^{\mu\nu}F_{\mu\nu}+\dfrac{\partial V(\phi)}{\partial\phi}-\bar{\psi}m_{q}\psi\dfrac{\partial G(\phi)}{\partial\phi}=0,
 \end{equation}

 \begin{equation}\label{13}
 -(i\gamma^{\mu}\partial_{\mu}+q\gamma^{\mu}A_{\mu})\psi+m_{q}G(\phi)\psi=0\qquad{\text{and}}\qquad
 \end{equation}
 
 \begin{equation}\label{15}
 \partial_{\mu}[G(\phi)F^{\mu\nu}]=-\bar{\psi}q\gamma^{\nu}\psi.
 \end{equation}
{where $\mu,\nu=0,1,2,3$. We are interested in studying static confining potentials that confine the electric flux generated in the model framework, leaving out the magnetic flux which will not be necessary for the current study. With this in mind, we choose magnetic field components $F_{ij}=0$, such that $F^{\mu\nu}F_{\mu\nu}=-F^{j0}F_{j0}=2E^2$ and the current density $j^\nu=\bar{\psi}q\gamma^{\nu}\psi$ restricted to charge density, i.e., $j^\nu=(\rho,\vec{0})$,  where $\rho=q/((4/3)\pi R^3)$.}
%we choose $\mu=k=1,\,2,\,3$ (spatial component) to ensure that only the static component of the scalar sector is available for the analysis. Since the objective is to analyse the {\it chromoelectric flux} confinement, we choose $\nu=0$ (time component), which implies that $j^k=0$, thus, the current density that is likely to generate {\it chromomagnetic flux} through Amp\`ere's law is eliminated. By these indices, the trace of the gauge fields leads to an electric field component $F^{\mu\nu}F_{\mu\nu}=-F^{j0}F_{j0}=2E^2$ and magnetic field component $F^{ij}F_{ij}=0$, with a charge density $j^\nu=\rho$, are sufficient for the analysis.} %We restrict the indices such that only the electric field is generated for the analyses, hence, $\nu=0$ and $\mu=j=1,2,3$ will be appropriate. 
%Notting that $\bar{\psi}q\gamma^{\nu}\psi=j^\nu=\rho$ where $\rho=q/((4/3)\pi R^3)$, %is the zeroth component of $j^\nu$ representing charge density. 
Thus, 
in spherical coordinates Eq.(\ref{15}) becomes
\begin{equation}\label{15a}
E_r=\dfrac{\lambda}{r^2G(\phi)},
\end{equation}
where $\lambda=q/4\pi\varepsilon_0$. Expanding Eq.(\ref{11}) in spherical coordinates, we obtain$^1$\footnotetext[1]{It is important to state at this point that the color dielectric function and the potential are related as $G(\phi)=(2\pi\alpha')^2V(\phi)$ \cite{Garousi}, where $\alpha'$ is the Regge slop with dimension of length squared. So it absorbs the dimension of the potential rendering the $G(\phi)$ dimensionless. But we set $(2\pi\alpha')=1$ throughout the paper for simplicity.} 
 \begin{equation}\label{23}
\dfrac{d^{2}\phi}{d r^{2}}+\dfrac{2}{r}\dfrac{d\phi}{d r}=\dfrac{\partial}{\partial\phi}\left[V(\phi)+\dfrac{\lambda^{2}}{2}\dfrac{1}{V(\phi)}\dfrac{1}{r^{4}}-\bar{\psi}m_{q}\psi V(\phi)\right] 
\end{equation}
we set $G(\phi)=(2\pi\alpha')^2V(\phi)$ and ignore the term in the order $\lambda^2$. Shifting the vacuum of the potential such that $\phi(r)\rightarrow \phi_0+\eta$, where $\eta(r)$ is a small perturbation about the vacuum $\phi_0=1/\alpha$. We can express Eq.(\ref{23}) as 
\begin{equation}\label{a2}
\nabla^2\eta=4\alpha^2(1-qm_{q}\delta(\vec{r}))\eta.
\end{equation}  
We can also determine the bosonic mass from the relation,
\begin{equation}\label{a3}
m^2_\phi=\dfrac{\partial^2V}{\partial\phi^2}|_{\phi_0}=4\alpha^2.
\end{equation}
%{\color{blue}this represents a pure gluon bound state.}
Also, we define the Dirac Delta function such that, 
\begin{equation}\label{27a}
  \delta(\vec{r})=
\begin{cases}
    \dfrac{1}{4\pi} \lim\limits_{R\to 0} \left(\dfrac{3}{R^{3}}\right), & \text{if}\,\; r\leq R\\
    0,              & r>R
\end{cases},
\end{equation}
where $R$ is the radius of the hadron and $r$ is the inter particle separation distance between the quarks. Now developing the Laplacian in Eq.(\ref{a2}) for particles inside the hadron gives
\begin{equation}\label{a4}
\nabla^2\eta=(m^2_\phi-qm_q^2)\eta,
\end{equation}
here, we have set 
\begin{equation}\label{a5}
R=\left(\dfrac{3m^2_\phi(2\pi\alpha')^2}{4m_q\pi} \right)^{1/3}.
\end{equation}
Similar expression can be seen in \cite{Sen,Boschi-Filho}, where the mass of the tachyonic mode is determined to be inversely related to the Regge slope i.e. $M^2\propto 1/\alpha'$ and $R\propto \sqrt{\alpha'}$. Accordingly, $m_\phi^2$ can be identified as the {scalar glueball mass obtained from bound state of gluons (quenched approximation) through the tachyon potential $V(\phi)$. Thus, the tachyonic field $\phi$ determines the dynamics of the gluons.} %and represents the bound state of the tachyons, has a proportionality relation with $m_q$. 
Expanding, Eq.(\ref{a4}) leads to
\begin{equation}\label{a6}
\eta''+\dfrac{2}{r}\eta'+K\eta=0,
\end{equation}
where $K=qm_q^2-m^2_\phi$. This equation has two separate solutions representing two different regimes i.e. the infrared (IR) and the ultraviolet (UV) regimes,
\begin{equation}\label{a7}
\eta(r)=\dfrac{\cosh(\sqrt{2|K|}r)}{\alpha r\sqrt{2|K|}}\qquad{\text{and}}\qquad \eta(r)=\dfrac{\sin(\sqrt{2K}r)}{\alpha r\sqrt{2K}} \qquad{\text{for}}\qquad r\,\leq\,R,
\end{equation}
respectively. We can now determine the confinement potential from the electromagnetic potential 
\begin{equation}\label{a8}
V_c(r)=\mp\int{E_r dr},
\end{equation}
where $E_r$ is the electric field modified by the {\it color dielectric function} $G(r)$. From Eq.(\ref{15a}) we can write,
\begin{equation}\label{a9}
E_r=\dfrac{\lambda}{r^2G(r)}.
\end{equation}
With the shift in the vacuum, $\phi\,\rightarrow\,\eta\,+\,\phi_0$, the tachyon potential and the color dielectric function takes the form
\begin{align}\label{a10}
V(\eta)=G(\eta)&=V(\phi)|_{\phi_0}+V'(\phi)|_{\phi_0}\eta+\dfrac{1}{2}V''(\phi)|_{\phi_0}\eta^2\nonumber\\
&=\dfrac{m^2_\phi}{2}\eta^2,
\end{align}
and the net confining potential becomes,
\begin{align}\label{a11}
V_c(r,m_q)=-\lambda\dfrac{\sqrt{(m^2_\phi+m_q^2)}\tanh\left[ \sqrt{2(m^2_\phi+m_q^2)}r\right]}{\sqrt{2}}.
\end{align}
Here, we choose the negative part of Eq.(\ref{a8}) corresponding to the potential of antiparticle, consequently, we choose $q=-1$ corresponding to an anti-color gluon charge. Besides, confinement in this regime increases with inter quark separation $r$ until a certain critical distance $r=r_*$, beyond $r_*$, hadronization sets in. %This is a clear indication that the quark and antiquark pair present here are light. 
Since $\lambda$ depends on the charge $q$, we can choose $\lambda=-1$. {Noting that $\sqrt{2(m^2_\phi+m_q^2)}r\ll 1$, the string tension can be determined from Eq.(\ref{a11}) as} 
\begin{align}\label{a112}
\sigma_c(m_q)&\simeq{m^2_\phi+m_q^2}\nonumber\\
&\simeq{m^2_L},
\end{align}
{where $m_L^2$ can be identified as the glueball-meson mixing state obtained from unquenched approximation \cite{Vento,Burakovsky,Genz,Simonov}.} Now, using $\sigma_c\sim 1 \text{GeV}/\text{fm}$ as adopted in \cite{Issifu,Issifu1} and $m^2_\phi=0.99\,\text{GeV}^2$ corresponding to $m_\phi=995\,\text{MeV}$, while $r_*=1/\sqrt{2\sigma_c}$ \cite{Bali1}. This glueball mass is closer to the scalar glueball mass determined for isoscalar resonance of $f_0(980)$ reported in the PDG as $(990\pm 20)\,\text{MeV}$ \citep{Tanabashi}. This choice of glueball mass was considered appropriate because we expect $m_q$ to be small but not negative or zero. This choice leads to $m_q=100\,\text{MeV}$, this is also within the range of strange (s) quark mass. In the PDG report $\bar{m}_s=(92.03\pm 0.88)\,\text{MeV}$ but, much higher values up to $400\,\text{MeV}$ are obtained using various fit approaches \cite{Sonnenschein} and $\bar{m}_s(1\,\text{GeV})=(159.50\pm 8.8)\text{MeV}$ are obtained using various sum rules \cite{Narison,Hatsuda}. In this model, the hadronization is expected to start when $r_*=1/\sqrt{2(m^2_\phi+m_q^2)}=0.71\,\text{fm}$, but for string models such as this, $r\,\gg\,r_*$ \cite{Bali1}.

On the other hand, the potential in the UV regime becomes
 \begin{align}\label{a12}
V_s(r,m_q)\simeq\left[-\dfrac{1}{2r}+\dfrac{(m_q^2-m^2_\phi)r}{3}+{\cal O}(r^3) \right] .
\end{align}
Here, we chose the positive part of the potential, i.e. a potential of a particle, so we set $\lambda=q=1$. Also with the string tension 
\begin{align}\label{a13}
\sigma_s&=\dfrac{m_q^2-m^2_\phi}{3}\nonumber\\
&=\dfrac{m^2_s}{3},
\end{align}
in both cases we have set the integration constants ($\tilde{c}$) to zero ($\tilde{c}=0$) for simplicity. Assuming that $\sigma_s\sim 1\,\text{GeV}/\text{fm}$ as used in \cite{Issifu}, we can determine $m_s=1.73\,\text{GeV}$, precisely the same as the lightest scalar glueball mass of quantum number $J^{PC}=0^{++}$ \cite{Tanabashi,Morningstar,Loan,Chen,Lee,Bali} and the critical distance is $r_{*s}=1/\sqrt{\sigma_s}$. The string tension $\sigma$ is related to the glueball mass by $m(0^{++})/\sqrt{\sigma}=\text{const.}$, where the magnitude of the constant is dependent on the approach adopted for the particular study. In this model framework, we have two different ratios, one in the IR regime and the other in the UV regime, i.e. $m_L/\sqrt{\sigma_c}=1$ and $m_s/\sqrt{\sigma_s}=1.73$ respectively. Hence, the ratio is dependant on the regime of interest. The string tension and the dimensionless ratio are precisely known physical quantities in QCD. Because their corrections are known to be the leading order of ${\cal O}(1/N_c^2)$ in $\text{SU}(\infty)$ limit. These quantities have been determined by lattice calculations, QCD sum rules, flux tube models and constituent glue models to be exact \cite{ Albanese,Bacilieri,Teper}. 

From the afore analyses, we can determine $m_q=2\,\text{GeV}$, while a bound state of quark-anti-quark pair would be $m_q^2=4\,\text{GeV}^2$. This is compared with the mass of a charm quark $\bar{m}_c(\bar{m}_c)=(1.280\pm 0.025)\,\text{GeV}$ as reported by PDG \cite{Tanabashi} and a slightly higher values up to $m_c=1.49\,\text{GeV}$ in various fits \cite{Sonnenschein}. Of cause, the model can take any quark mass greater than one ($m_q\,>\,1$) but less than some critical mass $m$. Hence, it could be used to successfully study chamonia and bottomonium properties. The only restriction here is that $m_\phi\,<\, m_q\,\leq\,m$ in the UV regime for the model to work efficiently. In fact, the major results in this section, i.e. the potentials and the string tensions, were reached in \cite{Issifu} and thoroughly discussed at $T=0$. So an interested reader can refer to it for step-by-step computations and justifications.

{In the model framework, we developed two different potentials, linear confining potential in the IR and a Cornell-like potential in the UV regimes. The potentials obtained from the UV and the IR regimes have confining strengths, $m_s^2$ and $m_L^2$ respectively. Linear confining models have been studied by several groups \cite{Eichten1,Quigg,Barnes1,Sauli,Leitao,Godfrey1,Deng} and its outcome establishes a good agreement with an experimental data for quarkonium spectroscopy together with its decay properties. This potential is motivated by string model for hadron, where quark and an antiquark pair are seen to be connected by a string that keeps them together. In spite of the successes of this model, it fails to explain the partonic structure of the color particles as observed from deep-inelastic scattering experiments. Consequently, since the QCD theory has generally been accepted as the fundamental theory governing strong interactions, the linear confining model gives a comprehensive description of some sector of the theory, identified as the IR regime \cite{Sundrum,Trawinski}. The Cornell potential on the other hand, calculated in the UV regime of the model is motivated by lattice QCD (LQCD) calculations \cite{Dudek,Meinel,Burch,Liu,McNeile,Daldrop,Kawanai,Burnier,Kalinowski}. Generally, it is represented as 
\begin{equation}
V(r)=-\dfrac{e}{r}+\sigma r,
\end{equation}
where $e$ and $\sigma$ are positive constants which are determined through fits. It consists of Coulomb-like part with strength $e=0.5$ due to single gluon exchange and the linear confining part with strength $\sigma=m_s^2/3$ in the model framework \cite{Bali1,Bali2,Alexandrou1}. It is a useful potential in QCD theory because it accounts for the two most important features of the theory i.e., color confinement and asymptotic free nature of the theory. It reproduces the quarkonium spectrum as well. Also, $e$ and $\sigma$ can be determined in the framework of heavy quark systems through fitting to an experimental data \cite{Eichten1,Lichtenberg}. 

Moreover, the IR model takes in quark masses in the range of $0$ to $1\, \text{GeV}$, while the UV model takes in masses greater than $\sqrt{3}$, consequently, there is an intermediate mass that can be exploited between $1$ and $\sqrt{3}\,\text{GeV}$. This mass range, is expected to show hadronization in the IR regime and an asymptotic free behaviour in the UV regime without a stable linear confining contribution (QGP). Likewise, the energy suitable for investigating the IR properties is $1.4\,\text{GeV}$ and below, deducing from the critical distance $r_*$, while in the UV regime the maximum energy for confinement is $1\,\text{GeV}$ and bellow judging from the magnitude of $r_{*s}$, but the process takes place at enormously high energy $r\rightarrow 0$, where Coulomb potential contribution is dominant. As a result, the model does not account for the behaviour of the particles at the intermediate energy regime, apart from knowing that the particles are 'asymptotically free' at $r\rightarrow 0$ and confined at $r\rightarrow r_{*s}$ and beyond. How the particles behave at the intermediate energy regimes is beyond the scope of this paper. Additionally, perturbative QCD (PQCD) involves hard scattering contribution which requires higher energy and momentum transfer in order to take place. This is referred to as the 'asymptotic limit', thus, the initial and final states of the processes are clearly distinct. Therefore, it has been widely argued that the current energy regime for exploring the QCD theory describes 'sub-asymptotic' regime, calling for a revision in pure perturbation treatment of the theory. In this case, new mechanisms or models are required to investigate the dynamics of elastic scattering in the {\it intermediate energy} region \cite{Coriano,Coriano1,Shifman,Nesterenko,Ioffe}. One way to investigate this energy region ('soft' behaviour) is to use dispersion relation together with Operator Product Expansion (quark-hadron duality) which are collectively referred to as  the QCD sum rule \cite{Coriano2,Reinders,Colangelo}. Thus, we can study the IR behaviour at the lower-end of the intermediate-energy region while the UV regime lies within the upper-end of the intermediate energy region to infinity. The CEBAF, Continuous Electron Beam Accelerator Facility, at Jefferson LAB was initially built to experimentally investigate these intermediate energy region before it was recently upgraded to cover the high energy regime.}

\section{Thermal Fluctuations}\label{TF}
In this section we discuss and propose how temperature can be introduced into the model. {Since we have determined in the previous section that the string tension which keeps the quark and antiquark pair in a confined state is dependant on the glueball-meson mixing states $m_L$ and $m_s$ \cite{Vento,Simonov}, we can determine the confining potentials at a finite temperature if we know exactly how the glueball-meson states fluctuate with temperature.} We calculate the temperature fluctuating {glueball-meson states} directly from Eq.(\ref{9}). Here, we will write the equations in terms of the glueball field $\eta$, redefined in dimensionless form such that $\chi=\eta/\phi_0$ for mathematical convenience. {The difference between the gluon field and the glueball fields $\eta$ are that, the glueball fields are massive whilst the gluon fields are not. In this paper, the field $\phi$ that describes the dynamics of the gluons {contains} tachyonic mode, when the tachyons condense they transform into glueballs $\eta$ with mass $m_\phi$.} Accordingly,
\begin{align}\label{tf1}
m^{*2}_\phi\phi_0^2=-\left\langle \dfrac{\partial^2\mathcal{L}}{\partial\chi^2} \right\rangle=\alpha^2\phi_0^2(2\pi\alpha')^2\langle F^{\mu\nu}F_{\mu\nu}\rangle +4\alpha^2\phi^2_0-4m_q(\alpha^2\phi_0^2)(2\pi\alpha')^2\langle\bar{\psi}\psi\rangle
\end{align}
{
%\begin{align}\label{tf1a}
%m_A^{*2}=-\left\langle \dfrac{\partial^2\mathcal{L}}{\partial F_{\mu\nu}^{i2}}\right\rangle &=\dfrac{1}{2}\langle G(\eta)\rangle\nonumber\\
%&=\dfrac{\phi_0^2m_\phi^2(2\pi\alpha')^2}{4}\langle\chi^2\rangle
%\end{align}
\cite{Carter} where the factor $(2\pi\alpha')^2$ was introduced to absorb the dimensionality of $V(\eta)$ and render $G(\eta)$ dimensionless as explained above. %That notwithstanding, it appears $m_A^{*2}$ is dimensionless in the above expression, which should not be the case. It should rather have the unit of mass squared, so to correct this, we rescale the gluon mass $m_A^{*2}\rightarrow m_A^{*2}/(2\pi\alpha')$, thus, Eq.(\ref{tf1a}) takes the form
%\begin{equation}\label{tf1b}
%m_A^{*2}=\dfrac{\phi_0^2m_\phi^2(2\pi\alpha')}{4}\langle\chi^2\rangle.
%\end{equation}
Besides, the theory of gluodynamics is classically invariant under conformal transformation giving rise to vanishing gluon condensation $\langle F_{\mu\nu}F^{\mu\nu}\rangle=0$ \cite{Issifu,Issifu1}. However, when the scale invariance is broken by introducing quantum correction, say $-|\epsilon_v|$, the gluon condensate becomes non vanishing i.e. $\langle F_{\mu\nu}F^{\mu\nu}\rangle\neq 0$. This phenomenon is referred to as the {\it QCD energy-momentum tensor trace ($\theta^\mu_\mu$) anomaly} \cite{Miller1}. In this paper, if one disregard the fermions and consider only the gluon dynamics in Eq.(\ref{9}), the potential $V(\phi)$ breaks the scale symmetry and brings about gluon condensation. This phenomenon bas been elaborated bellow in Sec.~\ref{QGC}.
} 

To determine the thermal fluctuations, we need to define the field quanta distribution of the mean gauge field $\langle F^{\mu\nu}F_{\mu\nu}\rangle$, $\langle\Delta^2\rangle$ with $\chi=\bar{\chi}+\Delta$ and the spinor fields $\langle\bar{\psi}\psi\rangle$. We find that the fluctuating scalar glueball mass $m^{*2}_\phi$ do not directly depend on the glueball field $\chi$ hence, no correction to the scalar field will be required. The field quanta distribution of the fields are$^2$, \footnotetext[2]{This follows the Matsubara formalism of field theory at finite temperature where space time becomes topological. Here, one makes use of the Euclidean imaginary time and solve the path integral by imposing a periodic condition on the gauge field, $\omega_n=2n\pi T$, and an antiperiodic condition, $\omega_n=(2n+1)\pi T$, for the fermion fields for $n\in \mathbb{Z}$. The imaginary time \cite{Landsman,Megias} transforms as $\beta=1/T$, where $T$ is temperature.} 
\begin{equation}\label{tf2}
\langle F^{\mu\nu}F_{\mu\nu}\rangle=-\dfrac{\nu}{2\pi^2}\int_0^\infty{\dfrac{p^4dp}{E_A}n_B(E_A)}\quad{\text{and}}\quad \langle\bar{\psi}\psi\rangle=\dfrac{\nu_q}{2\pi^2}\int_0^\infty{\dfrac{m_q p^2dp}{E_\psi}n_F(E_\psi)}
\end{equation} 
where $n_B(E_B)=(e^{\beta E_A}-1)^{-1}$ is the Bose-Einstein distribution function and $n_F(E_\psi)=(1+e^{\beta E_\psi})^{-1}$ is the Fermi-Dirac distribution function. 
%{ 
%In writing the first expression in the equation above, we observed that
%\begin{align}
%F^{\mu\nu}F_{\mu\nu}&=2\left(\partial^\mu A^\nu\partial_\mu A_\nu-\partial^\nu A^\mu \partial_\mu A_\nu \right) \nonumber\\
%&=-2A^\nu\left(\square g_{\mu\nu}-\left(1-\dfrac{1}{\alpha} \right)\partial_\mu\partial_\nu  \right)A^\mu,
%\end{align}
%in the second step we have subtracted the gauge fixing term $2/\alpha(\partial_\mu A^\mu)^2$. Choosing a Feynman gauge $\alpha=1$,
%\begin{align}
%F^{\mu\nu}F_{\mu\nu}&=-2A_\mu\square A^\mu.
%\end{align}
%}
%where in the last step we used $k\rightarrow i\partial_\mu$, $k$ is the momentum of the gluon.
Also, $\nu$ and $\nu_q$ are the degeneracies of the gluons and the quarks, respectively and $\beta=1/T$. % $k$ and $p$ are the momenta of the bosons and the fermions 
We can now analytically solve these integrals by imposing some few restrictions. We assume high energy limit, such that, $E\approx p$ for $c=1=\hbar$ corresponding to a single particle energy $E^2=m^{*2}+p^2$ therefore, we find \cite{Miller} 
\begin{align}\label{c16}
\langle F^{\mu\nu}F_{\mu\nu}\rangle&=-\dfrac{\nu}{2\pi^2}\int_0^\infty{dp \dfrac{p^4}{E_A} \dfrac{1}{e^{\beta E_A}-1} }\nonumber \\
&=-\dfrac{T^4\nu}{2\pi^2}\int_0^\infty{\dfrac{x^3dx}{e^{x}-1}}\nonumber\\
&=-\dfrac{\nu\pi T^4}{30}=-\dfrac{T^4}{T_{cA}^4}\langle F^{\mu\nu}F_{\mu\nu}\rangle_0,
\end{align}
which results$^3$ \footnotetext[3]{We used the standard integrals 
$$\int_0^\infty\dfrac{x^3dx}{e^x-1}=\dfrac{\pi^4}{15}$$
and 
$$\int_0^\infty\dfrac{xdx}{e^x-1}=\dfrac{\pi^2}{6}.$$
}
\begin{equation}\label{c16a}
\langle F^{\mu\nu}F_{\mu\nu}\rangle=\langle F^{\mu\nu}F_{\mu\nu}\rangle_0\left[1-\dfrac{T^4}{T^4_{cA}}\right],
\end{equation}
and from
\begin{align}\label{c18}
\langle\bar{\psi}\psi\rangle&=\dfrac{m_q\nu_q}{2\pi^2}\int_0^\infty{dp\dfrac{p^2}{E_\psi}n_F(E_\psi)}\nonumber\\
&=\dfrac{\nu_q m_q T^2}{2\pi^2}\int_0^{\infty}{\dfrac{xdx}{e^x+1}}\nonumber\\
&=\dfrac{T^2}{4T^2_{c\psi}}\langle\bar{\psi}\psi\rangle_0
\end{align}
 {we find the quark condensate}$^4$,\footnotetext[4]{Here, we used $$\int_0^\infty{\dfrac{xdx}{e^x+1}}=\dfrac{\pi^2}{12}.$$ 
}
\begin{equation}\label{c18a}
\langle\bar{\psi}\psi\rangle=-\dfrac{\langle\bar{\psi}\psi\rangle_0}{4}\left[1-\dfrac{T^2}{T_{c\psi}^2} \right] 
\end{equation}
Here, we used the transformation $x=E/T$ and the critical temperatures,
\begin{equation}\label{c19}
T_{cA}=\left( \dfrac{30\langle F^{\mu\nu}F_{\mu\nu}\rangle_0}{\nu\pi^2}\right)^{1/4} \qquad{\text{and}}\qquad T_{c\psi}=\left( \dfrac{6\langle\bar{\psi}\psi\rangle_0}{\nu_q m_q}\right)^{1/2}.
\end{equation}
However, from Eq.(\ref{tf1}) we can define the dimensionless quantity $\alpha^2\phi^2_0(2\pi\alpha')^2=\tilde{g}^2$, as a result,
\begin{align}\label{tf3}
m^{*2}_\phi\phi_0^2=\tilde{g}^2\langle F^{\mu\nu}F_{\mu\nu}\rangle +4\alpha^2\phi^2_0-4m_q \tilde{g}^2\langle\bar{\psi}\psi\rangle.
\end{align}
Now, substituting the thermal averages in Eqs.(\ref{c16a}) and (\ref{c18a}) of the fields into the above equation while we absorb the second term derived from the potential into the definition of the first and the third terms at their ground states, thus,
\begin{align}\label{tf4}
m^{*2}_\phi\phi^2_0=\tilde{g}^2\langle F^{\mu\nu}F_{\mu\nu}\rangle_0\left[1-\dfrac{T^4}{T_{cA}^4} \right] +\tilde{g}^2m_q\langle\bar{\psi}\psi\rangle_0\left[1-\dfrac{T^2}{T^2_{c\psi}} \right].
\end{align} 
To proceed, we use the standard definition for determining QCD vacuum,
\begin{equation}\label{tf5}
\theta^\mu_\mu=4V-\phi\dfrac{dV}{d\phi}.
\end{equation}
Using this expression and Eq.(\ref{a1}) we obtain
{
\begin{align}
\theta^\mu_\mu&=2[(\alpha\phi)^4-2(\alpha\phi)^2+1]-[2(\alpha\phi)^4-2(\alpha\phi)^2]\nonumber\\
&=2-2(\alpha\phi)^2,
\end{align}
maintaining the term with dependence on $\phi$ and discarding the constant, the vacuum becomes $B_0=m^2_\phi\phi^2_0/2$ \cite{Carter}.} Noting that $\langle F^{\mu\nu}F_{\mu\nu}\rangle_0=B_0$ represents the vacuum gluon condensate, we can express
\begin{equation}\label{tf6}
m^{*2}_\phi=m^2_\phi\left[1-\dfrac{T^4}{T_{cA}^4} \right]+m_q^2\left[1-\dfrac{T^2}{T^2_{c\psi}} \right],
\end{equation}
here, we set $\tilde{g}^2=2$, $\langle F^{\mu\nu}F_{\mu\nu}\rangle_0/\phi^2_0\approx m^2_\phi/2$ and $\langle\bar{\psi}\psi\rangle_0/\phi_0^2\approx m_q/2$. The string tension is determined by QCD lattice calculations to reduce sharply with $T$, vanishes at $T=T_c$ representing melting of hadrons \cite{Bicudo,Kaczmarek,Petreczky}.
At $T=0$ we retrieve the result for {glueball-meson mixing state $m^2_L$} in Eq.(\ref{a112}) i.e.,
\begin{equation}\label{tf7}
m^{*2}_\phi(0)=m^2_L=m_\phi^2+m^2_q.
\end{equation}
We can also retrieve the result in the UV regime $m^2_s$ if we set $m^2_\phi\rightarrow-m^2_\phi$ in Eq.(\ref{tf6}) at $T=0$. The nonperturbative feature of QGP is accompanied by change in the characteristics of the scalar or the isoscalar glueballs and the gluon condensate \cite{Kochelev}. This is evident in lattice QCD calculation of pure $\text{SU}(3)_c$ theory, pointing to sign changes \cite{Miller}. %The change in sign is necessary because the glueball mass in the UV regime is negative. 

Using Eqs.(\ref{a10}) and (\ref{tf6}), the {glueball potential} can be expressed as
\begin{align}\label{tf8a}
G(r,T)&=\left( \dfrac{\cosh[\sqrt{2}m_\phi^*r]}{m_\phi^*r}\right)^2\nonumber\\
&=\left( \dfrac{\cosh\left[\sqrt{2\left(m^2_\phi\left[1-\dfrac{T^4}{T^4_{cA}}\right] +m_q^2\left[1-\dfrac{T^2}{T^2_{c\psi}} \right] \right)}r\right] }{\sqrt{m^2_\phi\left[1-\dfrac{T^4}{T^4_{cA}}\right]+m_q^2\left[1-\dfrac{T^2}{T^2_{c\psi}} \right]}r}\right)^2.
\end{align}
Accordingly, we can express Eqs.(\ref{a11}), (\ref{a112}), (\ref{a12}) and (\ref{a13}) in terms of temperature as 
\begin{equation}\label{tf8}
V_c(r,T,m_q)=\dfrac{m_\phi^*\tanh[\sqrt{2}m_\phi^* r]}{\sqrt{2}},
\end{equation}
with string tension
\begin{equation}\label{tf9}
\sigma_c(m_q,T)={m^{*2}_\phi},
\end{equation}
and 
\begin{equation}\label{tf10}
V_s(r,m_q,T)=-\dfrac{1}{2r}+\dfrac{\left(m_q^2\left[1-\dfrac{T^2}{T^2_{c\psi}} \right]- m^2_\phi\left[1-\dfrac{T^4}{T_{cA}^4} \right]\right)r }{3},
\end{equation}
with string tension 
\begin{equation}\label{tf11}
\sigma_s(m_q,T)=\dfrac{\left(m_q^2\left[1-\dfrac{T^2}{T^2_{c\psi}} \right]- m^2_\phi\left[1-\dfrac{T^4}{T_{cA}^4} \right]\right)}{3}.
\end{equation}
As the temperature is increasing, the confining part shows some saturation near $T\approx T_{c\psi}\approx T_{cA}\approx T_c$ and vanishes completely at $T=T_c$ indicating the commencement of deconfinement and the initiation of QGP phase. This weakens the interaction of the particles such that the string tension that bind the $\bar{Q}$ and $Q$ reduces and eventually becomes asymptotically free at $T=T_c$. Beyond the critical temperature $T\,>\,T_c$ we have the QGP state where the quarks behave freely and in a disorderly manner \cite{Yagi,Pasechnik}. The possibility of studying the QGP state in detail with this model exists, but that is beyond the scope of this paper. {%\color{blue}Normalizing the string tension Eq.(\ref{tf9}) at $\sigma_0=m_q^2+m_\phi^2$ we can write,
When we set $\sigma_c(m_q,T)=0$ and calculate for a common critical temperature by assuming $T_{c\psi}=T_{cA}=1$ and $m^2_\phi=\sigma_c-m^2_q$ we will have
\begin{align}\label{phs}
&m^2_\phi\left[1-\dfrac{T^4}{T_{cA}^4} \right]+m_q^2\left[1-\dfrac{T^2}{T^2_{c\psi}} \right]=0\nonumber\\
&\sigma_c-(\sigma_c-m_q^2)T^4-m^2_qT^2=0,
\end{align}
where in the last step we substitute $m_\phi^2=\sigma_c(0)-m_q^2$, and in solving the equation we bear in mind that $\sigma_c\simeq 1\,\text{GeV}/\text{fm}$. Therefore,
\begin{equation}\label{phs1}
T_{c1}=\dfrac{1}{\sqrt{-1+m_q^2}},
\end{equation} 
consequently,
\begin{equation}\label{phs2}
\sigma_c(T)=m_\phi^2\left[1-\left(\dfrac{1}{\sqrt{-1+m_q^2}} \right)^4\dfrac{T^4}{T^4_{c1}}\right]+m_\phi^2\left[ 1-\left(\dfrac{1}{\sqrt{-1+m_q^2}}\right)^2\dfrac{T^2}{T^2_{c1}}\right].
\end{equation}
On the other hand, the common critical temperature in the UV regime can be calculated from,
\begin{align}\label{phs3}
&\dfrac{m_q^2-m_\phi^2}{3}-\dfrac{m^2_q T^2}{3}+\dfrac{m^2_\phi T^4}{3}=0\nonumber\\
&\sigma_s-\dfrac{m^2_q T^2}{3}+\dfrac{(m^2_q-3\sigma_s(0))T^4}{3}=0.
\end{align}
Substituting $\sigma_s\simeq 1\,\text{GeV}/\text{fm}$ and solving the equation,
\begin{equation}\label{phs3a}
 T_{c2}=\dfrac{\sqrt{3}}{\sqrt{-3+m_q^2}},
\end{equation}
accordingly,
\begin{equation}\label{phs4}
\sigma_s(T)=\dfrac{m_q^2\left[1-\left(\dfrac{\sqrt{3}}{\sqrt{-3+m_q^2}}\right)^2\dfrac{T^2}{T_{c2}^2} \right]-m_\phi^2\left[1-\left(\dfrac{\sqrt{3}}{\sqrt{-3+m_q^2}}\right)^4\dfrac{T^4}{T^4_{c2}}\right]  }{3}.
\end{equation}
}
%We present a phase diagram of $T_{c1}$ and $T_{c2}$ in  for the critical temperature corresponding to two regimes. Each critical temperature presents a different behaviour. 
%However, an analyses of these phenomena have been presented in Figs.~\ref{srg2} and \ref{srg1} but detail analyses of this subject is beyond the scope of this paper, we will deal with it in other papers in this series.}

It has been found that the bound state of charm-anti-charm state dissolve at $T=1.1T_c$ \cite{Lee1}. However, the question as to whether heavier quark bound states dissolve at $T=T_c$ \cite{Mocsy1,Gubler,Ding1} or temperatures higher than deconfinement temperature $T\,>\,T_c$ \cite{Asakawa,Datta,Ohno} still remain opened. These two separate pictures has informed different phenomenological models based on confinement and deconfinement transitions to QGP states to explain the observed suppression of $J/\psi$ produced in the RHIC. We present a simple model based on the projection that the $\bar{Q}Q$ bound state melt at $T=T_c$. %We can deduce from Eqs.(\ref{tf10}) and (\ref{tf11}) that the fermion mass plays a very important role in confining particles in the UV regime, following the restriction $m_q\,>\,m_\phi$. %To obtain a stable confinement in the UV regime $m_q$ must be bigger enough in order to compensate $T_{c\psi}$.

\section{Vector and Scalar Potentials, Gluon and Chiral Condensates}\label{VSG}
\subsection{Vector Potential}\label{VP}
To determine the vector potential, we solve Eq.(\ref{a4}) outside the hadron, i.e. $\delta(\vec{r})=0$, so
\begin{equation}\label{v1}
\eta''+\dfrac{2}{r}\eta'+K_0\eta=0,
\end{equation}
where $K_0=-m^2_\phi$. The solutions of this equation are 
\begin{equation}\label{v2}
\eta(r)=\dfrac{\cosh\left( \sqrt{2|K_0|}r\right) }{\alpha r\sqrt{2|K_0|}} \qquad{\text{and}}\qquad \eta(r)=\dfrac{\sin\left(\sqrt{2K_0}r\right)}{\alpha r\sqrt{2K_0}},
\end{equation}
for IR and UV regimes respectively, these solutions are equivalent to the solutions in Eq.(\ref{a7}) at $m_q=0$. Now, substituting the solution at the left side of Eq.(\ref{v2}) into (\ref{a9}) and into (\ref{a8}), we can determine the vector potential to be
\begin{align}\label{v3}
V_v(r)&=\mp \lambda\dfrac{m_\phi\tanh\left( \sqrt{2}m_\phi r\right)}{\sqrt{2}},
\end{align} 
with string tension 
\begin{equation}\label{v4}
\sigma_v\simeq\mp\lambda{m^2_\phi}.
\end{equation}
%in the second step of Eq.(\ref{v3}) we choose the negative part of the potential and $\lambda=-1$ for anticolor gluon charge corresponding to an antiparticle.
In terms of temperature fluctuations, the potential can be expressed as 
\begin{align}\label{v5}
V_v(T,r)=\mp\lambda\dfrac{m_{\phi A}^* \tanh\left(\sqrt{2}m_{\phi A}^* r\right)}{\sqrt{2}},
\end{align}
bearing in mind that,
\begin{equation}\label{v5a}
m^{*2}_{\phi A}=m^2_\phi\left[1-\dfrac{T^4}{T_{cA}^4}\right],
\end{equation}
the corresponding {temperature fluctuating string tension} \cite{Allen} becomes 
\begin{equation}\label{v6}
\sigma_v(T)\simeq\mp\lambda{m^{*2}_{\phi A}}.
\end{equation}
These results can also be derived from Eq.(\ref{tf6}) for $m_q=0$.

\subsection{Scalar Potential Energy}\label{SP}
The scalar potential energy \cite{Franklin,Castro,Crater} for confinement is calculated by comparing Eq.(\ref{13}) with the Dirac equation
\begin{equation}\label{tf12}
[c\hat{\alpha}\hat{p}+\hat{\beta}m_0c^2+S(r)]\psi=0,
\end{equation}
where $\hat{\alpha}$ and $\hat{\beta}$ are Dirac matrices and $S(r)$, which is well defined inside and on the surface of the hadron and zero otherwise is the scalar potential (for detailed explanations see \cite{Issifu}). Hence, the scalar potential obtained by comparing Eqs.(\ref{13}) and (\ref{tf12}) can be expressed as 
\begin{align}\label{tf13}
S(r)=m_qG(r)=2\alpha^2m_q\eta^2=\left[\dfrac{m_q}{(m^2_\phi-qm^2_q)r^2}\cosh^2\left( \sqrt{2(m^2_\phi-qm^2_q)} r\right)\right].
\end{align}
For an antiparticle we set $q=-1$, so we can rewrite the scalar potential in terms of the temperature as
\begin{equation}\label{tf14}
S(r,m_q,T)=\left[\dfrac{m_q}{m^{*2}_\phi r^2}\cosh^2\left( \sqrt{2}m_\phi^* r\right) \right].
\end{equation}
We can now write the net potential energy by adding the vector potential energy and the scalar potential energy, hence, 
\begin{equation}
V_{net}(r)=\mp\dfrac{q^2m_\phi\tanh\left(\sqrt{2}m_\phi r\right)}{4\sqrt{2}\pi}+\left[\dfrac{m_q}{m_\phi^2r^2}\cosh^2\left(\sqrt{2}m_\phi^*(0) r\right)\right],
\end{equation}
\cite{Dick,Cao} and in terms of temperature, we can express, 
\begin{equation}\label{s1}
V_{net}(r,m_q,T)=\mp\dfrac{q^2m_{\phi A}^* \tanh\left( \sqrt{2}m_{\phi A}^* r\right) }{4\sqrt{2}\pi}+\left[\dfrac{m_q}{m^{*2}_\phi r^2}\cosh^2\left(\sqrt{2}m_\phi^* r\right)\right].
\end{equation}

\subsection{Gluon Condensates}\label{QGC}
%\subsection{Gluon Condensate}
We calculate the energy momentum tensor trace, $\theta^\mu_\mu$, from the relation 
\begin{equation}\label{vg3a}
\theta_\mu^\mu=4V(\eta)+\eta\square\eta.
\end{equation}
Substituting the equation of motion Eq.(\ref{11}) into the above equation yields,
\begin{align}\label{vg3b}
\theta_\mu^\mu &=4V(\eta)-\eta\dfrac{\partial V}{\partial\eta}-\dfrac{\eta}{4}\dfrac{\partial G}{\partial\eta}F^{\mu\nu}F_{\mu\nu}+q\delta({\vec{r})}m_{q}\eta\dfrac{\partial G}{\partial \eta}\nonumber\\
&=4\tilde{V}+\tilde{G}F^{\mu\nu}F_{\mu\nu}-4q\delta({\vec{r}})m_{q}\tilde{G}\nonumber\\
&=4\tilde{V}_{eff}+\tilde{G}F_{\mu\nu}F^{\mu\nu},
\end{align}
where 
\begin{equation}\label{vg3c}
\tilde{V}=V-\dfrac{\eta}{4}\dfrac{\partial V}{\partial\eta},\qquad \tilde{G}=-\dfrac{\eta}{4}\dfrac{\partial G}{\partial\eta}\qquad{\text{and}}\qquad \tilde{V}_{eff}=\tilde{V}-q\delta(\vec{r})m_{q}\tilde{G}.
\end{equation}
{Moreover, the energy momentum tensor trace $\theta^\mu_\mu$ for the classical QCD chiral effective Lagrangian \cite{Pasechnik,Schaefer} is given as
\begin{equation}\label{vg3c1}
\theta^\mu_\mu=\sum\limits_f m_fq_f\bar{q}_f-\dfrac{b\alpha_s}{8\pi}F^{a\mu\nu}F^a_{\mu\nu},
\end{equation}
where $\beta(g)=-b\alpha_s/(4\pi)$ ($b=11$ depicts pure gluodynamics) is the QCD $\beta$-function, $m_f$ current quark mass matrix and $q_f$ is the quark field. %can be used to determine the running couplings. 
Simplifying $\tilde{V}_{eff}$ in Eq.(\ref{vg3b}) we find
\begin{align}\label{vg3c2}
\tilde{V}_{eff}(\eta)&=\dfrac{1}{4}m_\phi^2\eta^2-qm_q\delta(r)\left(-\dfrac{m_\phi^2}{4}\eta^2 \right) \nonumber\\
&=\dfrac{1}{4}m_L^2\eta^2 \qquad\text{for}\qquad q=1,
\end{align}
so
\begin{equation}\label{vg3c3}
\theta_\mu^\mu=\dfrac{1}{4}m_L^2\eta^2-\dfrac{G(\eta)}{2}F_{\mu\nu}F^{\mu\nu}\qquad\text{where}\qquad G(\eta)=V(\eta)=\dfrac{m_\phi^2\eta^2}{2}.
\end{equation}
Accordingly, comparing this with (\ref{vg3c1}), we can identify $\sum_fm_f=m^2_L$ (glueball-meson mixing mass), $\sum_fq_f\bar{q}_f=\eta^2$ (glueball field) and $b\alpha_s/(4\pi)=G(\eta)=-\beta(1/r^2)$ QCD $\beta$-function. To determine the strong running coupling, we note that $G(\eta)=\eta G'(\eta)/2$, so using the renormalization group theory \cite{Deur} 
\begin{equation}\label{vg3c4}
\beta(Q^2)=Q^2\dfrac{d\alpha_s(Q^2)}{dQ^2},
\end{equation}
we can deduce
\begin{equation}\label{vg3c5}
\beta(1/r^2)=-\eta\dfrac{d}{d\eta}\left(\dfrac{1}{2}G(\eta)\right), 
\end{equation}
consequently, the strong running coupling becomes $\alpha_s(1/r^2)=G(\eta)/2$, where we can relate $Q\rightarrow 1/r$. Using the solution at the right side of Eq.(\ref{a7}), we can express
\begin{align}\label{vg3c6}
\alpha_s(1/r^2)&=\left[1-\dfrac{2Kr^2}{3} \right] \nonumber\\
&=\left[1+\dfrac{2(m_\phi^2+m_q^2)r^2}{3} \right]\qquad{\text{for}}\qquad q=-1\rightarrow\nonumber\\
\alpha_s(q^2)&=\left[1-\dfrac{2}{3} \dfrac{m_L^2}{q^2}\right].
\end{align}
Notting that $Q^2\equiv-q^2$, where $Q^2$ is the space-like momentum and $q^2$ is the four-vector momentum. To eliminate the Landau ghost pole that occur at $q^2\rightarrow 0$ we assume that there is dynamically generated `gluon mass' at $q^2\rightarrow 0$ i.e. $q^2\cong m_A^2$ \cite{Zakharov,Furnstahl}.
%=)/+
}
Now relating the results in Eq.(\ref{vg3b}) with the standard vacuum expectation value of QCD energy-momentum tensor trace,
\begin{equation}\label{vc1}
\langle\theta^\mu_\mu\rangle=-4|\epsilon_v|,
\end{equation}
\cite{Issifu,Issifu1} we can write, 
\begin{equation}\label{vg5}
\langle \tilde{G}(\phi)F^{\mu\nu}F_{\mu\nu}\rangle=-4\langle |\epsilon_v|+\tilde{V}_{eff}(\eta)\rangle.
\end{equation}
Now we rescale $\tilde{V}_{eff}$ with the QCD vacuum energy density $-|\epsilon_v|$ such that, $\tilde{V}_{eff}\rightarrow -|\epsilon_v|\tilde{V}_{eff}$ and using the potential of the glueball fields defined in Eq.(\ref{a10}), we obtain 
\begin{equation}\label{vc2}
\left\langle\dfrac{m^2_\phi\eta^2}{4}F^{\mu\nu}F_{\mu\nu} \right\rangle=4|\epsilon_v|\left\langle  1-\dfrac{m^2_L\eta^2}{4}\right\rangle,
\end{equation}
here, we used $\tilde{V}_{eff}=1/4(m^2_\phi+qm_q^2)\eta^2=(m^2_L/4)\eta^2$ for $q=1$. Also, at the classical limit $\epsilon_v\rightarrow 0$ we have vanishing gluon condensate $\langle F_{\mu\nu}F^{\mu\nu}\rangle=0$. {The vacuum expectation value (VEV) of the gluon condensate was determined in \cite{Kondo,Cornwall} using Yang Mills theory with an auxiliary field $\phi$, where the fluctuations around $\phi$ gives rise to the glueball mass, $m(0^{++})$. It was observed that in the non-vanishing VEV, the gluon acquire a common mass, $m_A$, and the ratio $m(0^{++})/m_A$ in the leading order is 
\begin{equation}
\dfrac{m(0^{++})}{m_A}\cong\sqrt{6}.
\end{equation}
The non-vanishing condensate was attributed to the pair condensate of transverse gluons.
Juxtaposing that to the results in Eq.(\ref{vc2}) we can identify 
\begin{equation}\label{vc3}
m^2_A=\dfrac{m^2_L}{4}.
\end{equation}
}
{Other calculations based on QCD theory \cite{Gorbar} also shows that the non-vanishing gluon condensate in the absence of fermions can be expressed in terms of the gluon mass as 
\begin{equation}
m^2_g=\left(\dfrac{34N\pi^2}{9(N^2-1)} \langle\dfrac{\alpha_s}{\pi}G^{\mu\nu}G_{\mu\nu}\rangle\right)^{1/2}, 
\end{equation}
where $N$ is the number of colors and $\alpha_s$ is the strong coupling constant. This relation also show a proportionality between gluon mass and the non-vanishing condensate---more discussions relating to this subject can also be found in \cite{Fukuda,Kohyama}. Again, from the left side of Eq.(\ref{vc2}) we can deduce the expression $\langle(G(\eta)/2)F^{\mu\nu}F_{\mu\nu}\rangle=\langle\alpha_s(1/r^2)F^{\mu\nu}F_{\mu\nu}\rangle$ hence, the expression in the angle brackets at the right side of Eq.(\ref{vc2}) is equivalent to Eq.(\ref{vg3c6}).} %Since the gluon condensation is an IR property of QCD theory, the appearance of an IR mass $m^2_L=m^2_\phi+m^2_q$ which is identified in Eq.(\ref{a112}) as the string tension $\sigma_c$, in the vacuum Eg.(\ref{vc2}) is not a coincidence. 
Following the discussions at the latter part of Sec.~\ref{TM}, $m_L=1\,\text{GeV}$ so we can determine the gluon mass as $m_A=500\,\text{MeV}$. This can be compared to the result determined from QCD lattice simulation, projecting the gluon mass to be $m_A=600\sim 700\,\text{MeV}$ \cite{Leinweber,Langfeld,Alexandrou}. Some heavier gluon masses have also been determined closed to $\sim 1\,\text{GeV}$ using phenomenological analysis \cite{Field,Consoli} and some QCD lattice studies \cite{Kogan}. Quark and gluon condensates are responsible for confinement, glueball formation and hadron mass formation \cite{Shifman,Schaefer,Diakonov}. QCD vacuum at the ground state enable us to study the characteristics of the QGP, dynamics of phase transitions and hadronization. All these properties are as a results of nonperturbave nature of QCD theory in some regime (IR) and can not be studied using the usual perturbative QCD theory. 

To determine the temperature fluctuations in the gluon mass $m_A$ and the gluon condensate, we need to correct the temperature in the glueball field $\chi$. {The gluon mass is formed due to screening of the gluons in the vacuum or at a temperature where the constituent quarks dissolve into gluons \cite{Silva}.} The temperature can be introduced by defining the glueball field such that $\chi=\bar{\chi}+\Delta$, so $\eta=\phi_0(\bar{\chi}+\Delta)$, with the restriction $\langle\Delta\rangle=0$ to avoid the occurrence of cross terms in the thermal average.  Again, $\bar{\chi}$ represents the mean glueball field and the angle brackets represent thermal average. We express the fluctuation $\langle\Delta^2\rangle$ in terms of field quanta distribution as defined for the gauge and the fermion fields in Eqs.(\ref{c16a}) and (\ref{c18}) respectively i.e.
\begin{align}\label{gb6}
\langle\Delta^2\rangle&=\dfrac{1}{2\phi_0^2\pi^2}\int_0^\infty{dk \dfrac{k^2}{E_\eta}\dfrac{1}{e^{\beta E_\eta}-1}}\nonumber\\
&=\dfrac{T^2}{2\phi_0^2\pi^2}\int_0^\infty{\dfrac{xdx}{e^x-1}}\nonumber\\
&=\dfrac{T^2}{12\phi_0^2}.
\end{align}
Reverting to Eq.(\ref{11}) and the definition $\eta=\phi_0(\bar{\chi}+\Delta)$ we can determine the thermal average directly from the equation of motion to be 
\begin{align}\label{gb6a}
&\left\langle \dfrac{1}{4}\dfrac{\partial G}{\partial\eta}F^{\mu\nu}F_{\mu\nu}\right\rangle =\left\langle\dfrac{\partial V(\eta)}{\partial\eta}(qm_q\delta(r)-1) \right\rangle \nonumber\\
&\left\langle \dfrac{m^2_\phi\eta}{4} F^{\mu\nu}F_{\mu\nu}\right\rangle=-\langle (m^2_\phi-qm_q^2)\eta\rangle, 
\end{align}
clearly $\bar{\eta}=\bar{\chi}=0$ is a solution, hence, $\langle\eta^2\rangle=\phi_0^2\langle\Delta^2\rangle$.

Moreover, substituting Eq.(\ref{gb6}) for temperature fluctuations in $\eta^2$ into Eq.(\ref{vc2}), the gluon condensate becomes 
{
\begin{align}\label{vc4}
\left\langle\dfrac{m^2_\phi\eta^2}{4}F^{\mu\nu}F_{\mu\nu} \right\rangle&=4|\epsilon_v|\left[1-\dfrac{m_L^2\phi_0^2}{4}\left(\dfrac{T^2}{12\phi_0^2}\right)  \right] \nonumber\\
&=4|\epsilon_v|\left[1-\dfrac{T^2}{T^2_{c\eta}}\right]\qquad{\text{where}}\qquad T^2_{c\eta}=\dfrac{48}{m_L^2}.
\end{align}
}
The QCD vacuum at an extremely high temperature and density has an important significance in both elementary particle physics and cosmology. %The vacuum consists of bound states of quarks $m^2_q$ and gluons $m_\phi^2$, i.e. a hybrid vacuum. 
In cosmology, it can be used to explain the evolution of the universe at extremely higher temperatures when matter becomes super dense and hadrons dissolve into a `soup' of their constituents --- quarks and gluons. At such superdense matter regimes the quarks become very close and asymptotically free so, the quarks are no longer confined into hadrons. It is important to note that the energy density is related to temperature as $\rho\propto T^4$, so the density increases with temperature. This phase of matter was proposed \cite{Collins,Cabibbo} after the asymptotic free \cite{Gross,Politzer} nature of the QCD theory \cite{Fritzsch} was determined. Currently, almost all the accepted models for the cooling of the universe are based on phase transitions, either first or second order of spontaneous symmetry breaking of fundamental interactions \cite{Linde,Bailin,Boyanovsky}. On the other hand, the standard model (SM) of elementary particle physics suggests two of such phase transitions \cite{Boyanovsky}. It shows an electroweak symmetry (EW) breaking at temperatures in the order of $T\sim 100\,\text{GeV}$ which generates mass to elementary particles such as quarks. It is also associated with the observed baryon-number-violation of the universe \cite{Trodden}. This transition is determined to be an analytical crossover in lattice simulations \cite{Kajantie1}. 

Again, spontaneous chiral symmetry breaking is next, it is known to occur at temperatures in the order of $T\sim 200\,\text{GeV}$. Nevertheless, we have hadronic matter bellow this temperature and above it we have an expected transit into the QGP phase. This phase is also important in understanding the evolution of the early universe. Some references on this subject using nonperturbative lattice QCD models can be found in \cite{Susskind,Petreczky1}. Particularly, knowing that the baryon chemical potential $\mu_B$ is expected to be smaller than the usual hadron mass, $\mu_B\approx 45\,\text{MeV}$ at $\sqrt{s_{NN}}=200\,\text{GeV}$ \cite{Adams}, and almost vanishing $\mu_B\approx 0\,\text{GeV}$ in the early universe. It is suitable to model inflation of the early universe at high temperatures and low baryon densities. %This model is based on $\mu_B=0\,\text{GeV}$ which is in line with many QCD lattice calculations. 
There is strong evidence that confinement of quarks into hadrons is a low energy phenomena \cite{Susskind,Aoki,Bhattacharya}, but it strongly suggests that QCD phase transition is a crossover. Numerically, it has been established that at $\mu_B=0\,\text{GeV}$ the two phase transitions possibly coincide i.e., deconfinement and chiral symmetry restoration rendering the chiral effective theory invalid \cite{Cheng}. From our model framework, Eq.(\ref{vc4}), we obtain confinement at $T\,<\,T_c$, deconfinement and restoration of chiral symmetry at $T=T_c$ and QGP phase at $T\,>\,T_c$.

Deducing from Eq.(\ref{vc4}) the thermal fluctuating gluon mass becomes
{
\begin{align}\label{gb6b}
m^{*2}_A&=\left(\dfrac{m^2_L}{48\phi_0^2} \right) T^2\nonumber\\
&=g^2T^2
\end{align}
{where $g^2=m^2_L/48\phi^2_0$} is a dimensionless coupling constant \cite{Blaizot,Silva}}. This results looks similar to the Debye mass obtained from the leading order of QCD coupling expansion at higher temperatures and zero chemical potential, $m_D=gT$ \cite{Kajantie}. This mass is obtained from the lowest order perturbation QCD theory \cite{Nadkarni} but it is known to be resulting from the IR behavior of the theory \cite{Manousakis}. This mass occur at temperatures greater than the deconfinement temperature $T_c$ due to the melting of $\bar{Q}Q$ bound states \cite{Matsui,Kajantie} generally, it is referred to as {\it Debye screening mass of color charges}.

{Furthermore, the experimental evidence of QCD vacuum condensates dates back from the pre-QCD era $1950-1973$ to post-QCD after $1974$ and it still remains an issue of interest. The QCD vacuum condensate informs color confinement and chiral symmetry breaking which are fundamentals of strong interactions. The condensate is a very dense nonperturbative state of matter consisting non-vanishing gluon and quark condensates, in theories containing effective quarks, interacting in a haphazard manner \cite{Shifman}. Since quarks and gluons are not observed directly in nature, these characteristics are difficult to observe experimentally, only colorless (baryons) or color neutral (mesons) hadrons are observed. However, much is known about the QCD vacuum, credit to fast evolving QCD sum rule which explore the gauge invariant formalism to explain the nonperturbative nature of the condensates \cite{Kuzmenko,Simonov1}. That notwithstanding, the QCD sum rule gives good results at the intermediate energy region and poses some short falls at large distance regime where color confinement and chiral symmetry breaking are more pronounced. So a resort is made to Vacuum Correlator Method (VCM) to give a comprehensive description of all the possible QCD phenomena involved \cite{Dosch,Simonov2,Giacomo}. The VCM is based on gauge invariant Green's function of colorless or color neutral objects expressed in path integral formalism using field correlators instead of propagators.}

\subsection{Chiral Condensate}\label{CC}
{\it Chiral symmetry breaking} is one of the known physical properties of the non perturbation regime of QCD theory aside the famous {\it color confinement}. Mass splitting of chiral partner observed in hadron spectrum and {\it Goldstone bosons} ($\pi^{\pm}\,\text{and}\,\pi^0$) which appear due to spontaneous symmetry breaking leading to {\it color confinement} are strong evidence of chiral symmetry breaking \cite{Bicudo1,Suganuma,Kitano} in QCD vacuum \cite{Weinberg}. {The presence of quark mass breaks the chiral symmetry explicitly.} With these background and other theoretical considerations, it is believed that there is chiral condensate in the vacuum which is proportional to the expectation value of the fermionic operator or {the quark condensate $\langle\bar{\psi}\psi\rangle$.} However, increasing temperature augments the thermal excitations of the hadrons due to an increase in the density ($\rho\propto T^4$) of their quark constituents. This decreases the vacuum condensate until it eventually vanishes at a critical temperature $T_{cq}$ resulting into restoration of the chiral symmetry. At this temperature the hadrons undergo phase transition to deconfinement and quark-gluon plasma (QGP) phase. Increasing baryon density have the same consequences on the QGP phase and chiral condensate similar to temperature increase \cite{Borsanyi,Bazavov}. To begin the calculation of the {\it chiral condensate} one needs to take into account the dependence of the hadron masses on the current quark mass $m_q$. Some known first principle approach to this subject that give a consistent account of hadrons and their quark mass dependant are; {\it Chiral Perturbation Theory} (ChPT) \cite{Borasoy}, lattice QCD (lQCD) \cite{Durr} and {\it Dyson-Schwinger Equation} (DSE) \cite{Holl,Flambaum}.

However, the dominant degrees of freedom for QCD at low energies are hadrons \cite{Leutwyler, Verbaarschot}. Particularly, pions and kaons which interact weakly according to Goldstone theory, hence, they can be treated as free particles. Consequently, the standard relation for calculating the chiral condensate, $\langle\bar{q}q\rangle$, can be derived from Eq.(\ref{9}) as
\begin{equation}\label{cc1}
\langle\bar{q}q\rangle=-\left\langle\dfrac{\partial\mathcal{L}}{\partial m_q} \right\rangle, 
\end{equation}
\cite{Borasoy} where $\langle\bar{q}q\rangle$ is significant for {\it spontaneous chiral symmetry breaking} (S$\chi$SB) and the angle brackets represent thermal average. %The negative sign was introduced to represent the thermal fluctuations of the {\it pion gas}, that notwithstanding, a conserve quantity remains conserved even when multiplied by a constant. 
{Subtracting the fundamental vacuum condensate $\langle\bar{q}q\rangle_0$ to ensure that the condensate has its maximum $T=0$ \cite{Koch}, we can express the above relation as} 
\begin{equation}\label{cc2}
\langle\bar{q}q\rangle=\langle\bar{q}q\rangle_0-\left\langle\dfrac{\partial\mathcal{L}}{\partial m_q} \right\rangle,
\end{equation}
here, details of the vacuum condensate $\langle\bar{q}q\rangle_0$ \cite{Jankowski,Dashen} is of no importance to the analyses. All the information needed to study the contributions of the hadrons are contained in the medium dependant term. From Eq.(\ref{9}) and using the expression in Eq.(\ref{a10}) derived from $\phi(r)\rightarrow\phi_0+\eta$, we obtain 
\begin{align}\label{cc3}
\left\langle\dfrac{\partial\mathcal{L}}{\partial m_q} \right\rangle&=\langle G(\eta)\bar{\psi}\psi\rangle \nonumber\\
&=\left\langle \dfrac{m^2_\phi\eta^2}{2}\bar{\psi}\psi \right\rangle .
\end{align}  
Thus, the chiral condensate is proportional to the {glueball potential} $G(\eta)$ which has its maximum condensate at $G(\eta)\rightarrow 0$ \cite{Issifu,Issifu1}. The quark density $\langle\bar{\psi}\psi\rangle$ is a measure of the strength of the condensate and S$\chi$SB. At $\langle\bar{\psi}\psi\rangle=0$ we have an exact chiral symmetry, deconfinement and QGP phase, whilst nonvanishing quark condensate, $\langle\bar{\psi}\psi\rangle\neq 0$, is the regime with S$\chi$SB and {\it color confinement} \cite{Hoofta,Coleman}. Therefore, $\langle\bar{\psi}\psi\rangle$ serves as an order parameter that determines the phase transitions. Since the explicit pion degrees of freedom is significant for studying chiral symmetry restoration at low temperature and quark densities, $\langle\bar{\psi}\psi\rangle$ is constituted with {\it up} (u) and {\it down} (d)  quarks. Eventually, protons and neutrons are also composed of similar quark constituents hence, in QCD with two flavour constituents, 
\begin{equation}\label{cc3a}
\langle\bar{\psi}\psi\rangle=\langle\bar{q}_iq_i\rangle=\langle\bar{u}u\rangle+\langle\bar{d}d\rangle.
\end{equation}
Additionally, in terms of temperature
\begin{align}\label{cc4}
\langle\bar{q}q\rangle&=\langle\bar{q}q\rangle_0-\left[\dfrac{m_\phi^2}{2}\left(\dfrac{T^2}{12} \right)\left(\dfrac{m_q\nu_q T^2}{24} \right)   \right]\nonumber\\
&=\langle\bar{q}q\rangle_0-\left[\dfrac{m_\phi^2m_q\nu_q}{576}T^4 \right]\nonumber\\ 
&=\langle\bar{q}q\rangle_0\left[1-\dfrac{T^4}{T_{cq}^4}\right],
\end{align}
where in the first step, we used the results in Eqs.(\ref{c18}) and (\ref{gb6}), bearing in mind that $\langle\eta^2\rangle\simeq\phi_0^2\langle\Delta^2\rangle$ and
\begin{equation}\label{cc5}
T_{cq}=\left(\dfrac{576}{m_\phi^2m_q\nu_q}\langle\bar{q}q\rangle_0\right)^{1/4}.
\end{equation}
Thus, at $T=T_{cq}$ the chiral symmetry gets restored in the model framework.
{If one intends to investigate the behavior of the chiral condensation with varying quark mass $m_q$, we can define the critical temperature as 
\begin{equation}\label{cc5a}
T_{cq}=\left(\dfrac{576}{m_\phi^2\nu_q}\langle\bar{q}q\rangle_0\right)^{1/4},
\end{equation}
corresponding to a condensate
\begin{equation}\label{cc5b}
\langle\bar{q}q\rangle=\langle\bar{q}q\rangle_0\left[1-m_q\dfrac{T^4}{T_{cq}^4}\right].
\end{equation}
} 
Also, an evidence from lQCD confirm that at the confinement phase the chiral symmetry is spontaneously broken down \cite{Coleman,Fiorilla} to flavour group i.e.
\begin{equation}\label{cc6}
\text{SU}(2)_R\times\text{SU}(2)_L\rightarrow \text{SU}(2)_V,
\end{equation}
with associated three Goldstone bosons ($\pi^{\pm}\,\text{and}\,\pi^0$) which spontaneously break the chiral symmetry. For three quark flavors we have 
\begin{equation}\label{cc7}
\text{SU}(3)_R\times\text{SU}(3)_L\rightarrow \text{SU}(3)_V,
\end{equation}
here, there are eight Goldstone bosons ($\pi^{\pm},\,\pi^0,\, K^{\pm},\,K^0,\,\bar{K}^0\,\text{and}\,\eta$) involved ({see references \cite{Lenaghan,Collins1})}.

{Again, a highly excited states in high energy hadronic collisions leads to the formation of disoriented chiral condensate which can latter decay into ordinary vacuum through coherent emission of low momentum pions. This process is theoretically synonymous to the Higgs mechanism that leads to the release of Goldstone bosons. This leaves a signature of color confinement eventually \cite{Mohanty}. Even though the idea of chiral condensate was speculative at its inception in 1990's \cite{Bjorken,Blaizot1,Nelson,Anselm} it has attracted several theoretical and experimental attention subsequently. Aside its simplicity, it is also motivated by the discovery of the 'so called' Centauro events in cosmic-ray \cite{Lattes,Augusto,Gladysz} where clusters consisting charged and neutral pions were observed. The chiral condensate has also been studied theoretically in the light of high energy heavy ion collisions \cite{Rajagopal} due to the high energies involved at the collision zone, a hot chirally symmetric state (QGP) is formed in the process. Because of the fast expanding nature of the system at the early stages, it is quenched down to a low temperature where chiral symmetry is spontaneously broken down. Several experiments have since been set up to investigate this phenomena, key among them  are the cosmic-ray experiment \cite{Lattes,Augusto}, nucleon-nucleon collisions at CERN \cite{Arnison,Alpgard,Alner}, Fermi LAB \cite{Melese} particularly, MiniMAX experiment \cite{Brooks}, nucleon-nucleon collisions at CERN SPS \cite{Aggarwal,Appelshauser}, and the RHIC \cite{Nakamura,Ackermann}. It also forms part of the heavy ion collision programme carried out with the multi purpose detector ALICE, at LHC \cite{Collaboration,Angelis}.}

\section{Conclusion}\label{CON}
%{\color{blue}
%\subsection{Analysis}

% at  In this regard, the heavy quark mass serves as a cut-off, so the partonic hard scattering process can be investigated in the framework of nonperturbative QCD theory due to low transverse momentum. As a results, proper understanding of heavy flavour production and comparison with experimental data will enhance the existing understanding and a testing medium for both perturbative and nonperturbative QCD calculations.  

%The experimental programmes at SPS and RHIC for heavy ion collisions  were aided by the LHC programme due to large magnitude of center-of-mass energy $\sqrt{s}$ between $2.76$ to $8\,\text{TeV}$. This has enriched the investigation of heavy-flavour production in terms of  magnitude, new observables and precision. A detailed knowledge on the heavy-flavour production will prepare the grounds investigating some high interest scalar particles such as, weak boson production, Higgs boson production and physics beyond the standard model.
%\subsection{Conclusion}
Following the discussions in \cite{Issifu} for the constituent quark masses, we can deduce that the constituent quark masses of this model are $M(r\rightarrow r_*)=2m_q=200\,\text{MeV}$ and $M(r\rightarrow 0)=2m_q=4\,\text{GeV}$ for the IR and the UV regimes respectively. Thus, the potential in the IR and the UV regimes can take masses within the ranges $0\,\leq\,m_q\,\leq 200\,\text{MeV}$ and $2\,\leq\,m_q\leq\,4\,\text{GeV}$ respectively. Hadronization is expected to set in, in the IR and the UV regimes for $m_q\,>\,200$ and $m_q\,>\,4$ respectively. Since we have adopted the lattice simulation results for the string tension $\sigma\sim 1\,\text{GeV}/\text{fm}$, and noticing that $\sigma=1/(2\pi\alpha')$ as shown in many confining string models \cite{Boschi-Filho,Boschi-Filho1}. The choice $(2\pi\alpha')^2=1$ used throughout the paper is in order. In the framework of the model, the glueball field $\chi$ do not contribute to the fluctuations in the scalar glueball mass $m^{*2}_\phi$. While the only candidate that contribute to the fluctuating gluon mass $m_A^{*2}$ is the glueball field. On the other hand, the gauge fields, the spinor fields and the glueball fields, all contribute to the gluon condensate $\langle(m^2_\phi\eta^2/4)F^{\mu\nu}F_{\mu\nu}\rangle$ and chiral condensate. Thus, the condensates are important in understanding QCD theory but difficult to study experimentally due to the haphazard nature of interactions among these fields in the vacuum. Also, $\nu_q$ and $\nu$ are the degeneracies of quarks and gluons and they take in the values, $6$ and $16$ for $\text{SU}(2)$ and $\text{SU}(3)$ representations respectively. These degeneracies are important in determining the critical temperatures of the model. The critical temperatures are small when the degeneracies are infinitely large and when there is no degeneracies at all ($\nu,\,\nu_q\rightarrow 0$) the critical temperature becomes infinitely large, same is true for quarks and gluons as presented in Eqs.(\ref{cc6}) and (\ref{c19}).

The model produces two forms of temperature corrections to the string tension, $-T^2$ coming from the spinors (quarks) and $-T^4$ from the gauge fields (gluons). The $-T^2$ correction to the string tension has been corroborated by some QCD lattice calculations \cite{Kaczmarek,Pisarski,Forcrand} and some phenomenological models \cite{Issifu1}. That notwithstanding, $-T^4$ correction to the string tension has also been proposed by some phenomenological models \cite{Boschi-Filho}. Both corrections give the correct behavior of the string tension, i.e. it should reduce sharply with temperature and break or vanish at $T=T_c$ for simple models such as the one discussed here. For simplicity, we used $T_{cA}=T_{c\psi}$ for some of the analyses ---particularly, the potentials--- but there is no evidence that these two critical temperatures have the same magnitude. In any case, such assumption is informed and does not affect the results or the analyses. However, there is a discussion in \cite{Fukushima} suggesting that $T_{c\psi}\,>\,T_{cA}$ ($T_{c\psi}\sim 270\,\text{MeV}$ and $T_{cA}\sim 170\,\text{MeV}$), or at least a discrepancy of about $3\%$ reported in \cite{Cheng1}. Using the magnitude of the string tension and the scalar glueball mass $m_\phi$ calculated above, we obtained two different glueball-meson states corresponding to $m_L=1\,\text{GeV}$ and $m_s=1.73\,\text{GeV}$ for the IR and UV regimes respectively. The gluon mass was also determined as $m_A=500\,\text{MeV}$. The critical distance for confinement in the IR regime has been determined to be $r_*=0.71\,\text{fm}$, and its corresponding value in the UV regime is $r_{*s}=1/\sqrt{\sigma_s}=1\,\text{fm}$. Similarly, $r_*$ and $r_{*s}$ can be expressed as a function of temperature like the string tensions.

\begin{figure}[H]
  \centering
  \subfloat[Left Panel]{\includegraphics[scale=0.6]{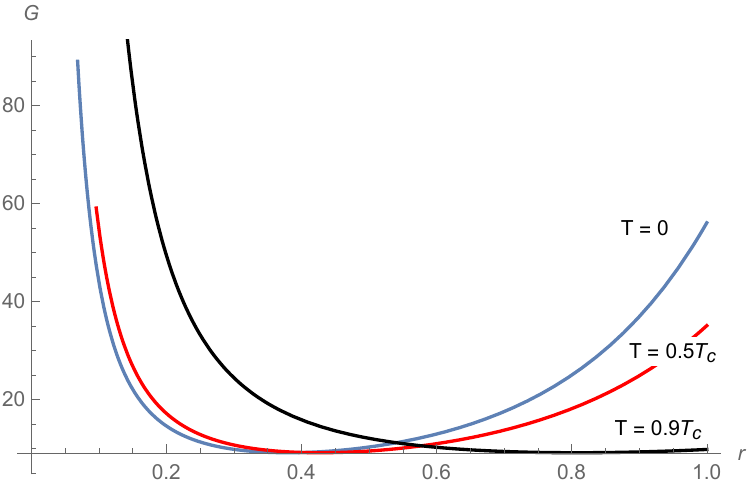}}
  \qquad
  \subfloat[Right Panel]{\includegraphics[scale=0.6]{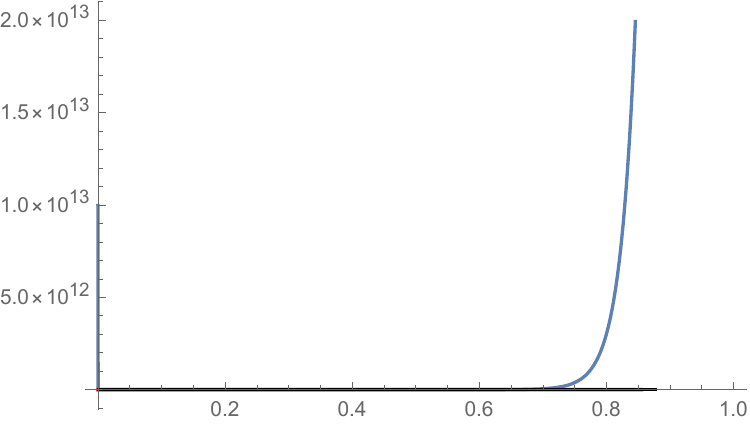}}
 \caption{A graph of glueball potential $G(r,T)$, against $r,T$ for $m_q=0.1$ (left) and $m_q\rightarrow\infty$ (right).}
   \label{pt1}
   \floatfoot{The glueball condensation increases with increase in depth of the curve. Hence the condensate increases from $T=0.9T_c$(black), $T=0.5$(red) to $T=0$(blue). %\cite{Pasechnik,Brambilla}. Within this temperature range, there is confinement and chiral symmetry breaking. In the right panel, we show how increase in $m_q\rightarrow \infty$ affects the behaviour of the potential and confinement of the particles. As $m_q$ is increasing, the potential increases and the confinement becomes stronger. %If we continue increasing $m_q$ beyond the threshold or the critical mass $m$, the black curve ($T=0.9T_c$) will vanish followed by the vanishing of the red curve ($T=0.5T_c$). And at $T=T_c$ the potential simply vanishes showing deconfinement phase. %For $T\,>\,T_c$ we move into the QGP phase. The only curve that survives under such extreme mass condition is the blue curve ($T=0$), but gets flat at the top showing hadronization \cite{Issifu}. This means that, there is a mass limit within which this model can function efficiently with temperature variation.
   }
\end{figure}

\begin{figure}[H]
  \centering
  \subfloat[Left Panel]{\includegraphics[scale=0.6]{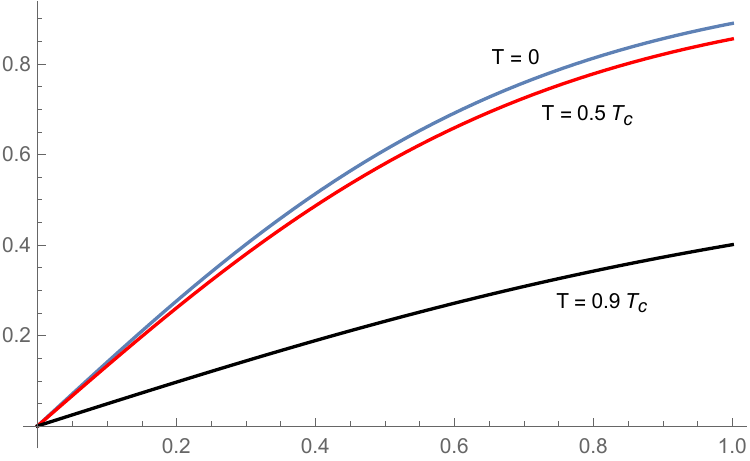}}
  \qquad
  \subfloat[Right Panel]{\includegraphics[scale=0.6]{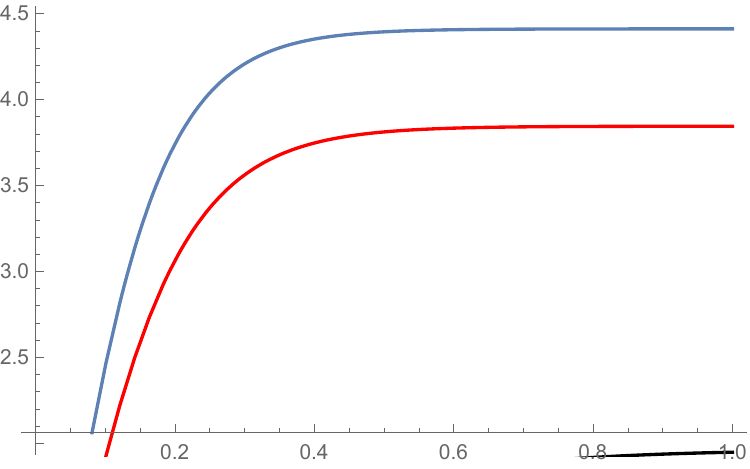}}
 \caption{A graph of net confining potential, $V_{c}(r,T)$, against $r,T$ for $m_q=0.1$ (left) and $m_q\rightarrow\infty$ (right).}
   \label{pt}
   \floatfoot{As it is shown on the left panel, the gradient of the graph increases with decreasing temperature from $T=0.9T_c$(black), $T=0.5$(red) to $T=0$(blue) \cite{Pasechnik,Brambilla}. Within this temperature range, there is confinement and chiral symmetry breaking. In the right panel, we show how increase in $m_q\rightarrow \infty$ affects the behavior of the potential and confinement of the particles. As $m_q$ is increasing, the potential increases and the confinement becomes stronger. %If we continue increasing $m_q$ beyond the threshold or the critical mass $m$, the black curve ($T=0.9T_c$) will vanish followed by the vanishing of the red curve ($T=0.5T_c$). And at $T=T_c$ the potential simply vanishes showing deconfinement phase. %For $T\,>\,T_c$ we move into the QGP phase. The only curve that survives under such extreme mass condition is the blue curve ($T=0$), but gets flat at the top showing hadronization \cite{Issifu}. This means that, there is a mass limit within which this model can function efficiently with temperature variation.
   }
\end{figure}

\begin{figure}[H]
  \centering
  \subfloat[Left Panel]{\includegraphics[scale=0.6]{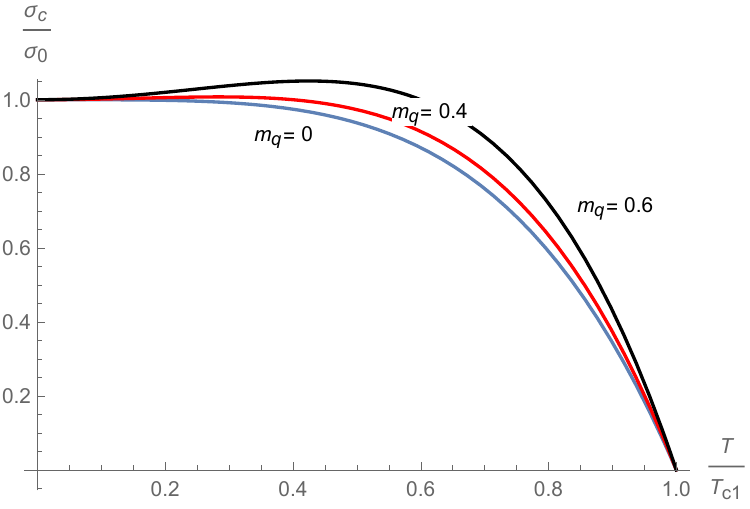}}
  \qquad
  \subfloat[Right Panel]{\includegraphics[scale=0.6]{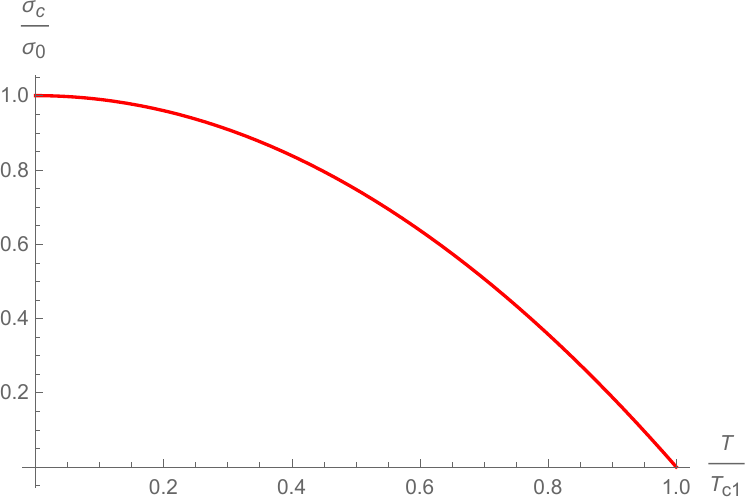}}
  \caption{A graph of string tension, $\sigma_c(T)/\sigma_0$, against $T/T_{c1}$ for different values of $m_q$ (left) and $m_q\rightarrow\infty$ (right). }
   \label{st}
   \floatfoot{Here, we show the behavior of the string tension in the IR regime with varying temperature. The string breaks quickly for light quarks while the heavier quarks have relatively longer life time even though they all vanish at $T=T_{c1}$. A regime where all the bond states are expected to dissolve into a `soup' of their constituents. It also gives an insight into the behavior of the string tension for chiral limit $m_q=0$ and chiral effective $m_q\neq 0$ regimes. %The string tension also increases with increasing $m_q$ as displayed in the right panel. %We find that the string decreases faster with increasing temperature. The graph in the left panel, is sketched for $m_q=0.1$ and $m_\phi=1$ and that in the right panel is for $m_q\rightarrow\infty$. Indeed, an increase in $m_q$ goes into increasing the string tension. %In all these cases the string tension breaks at $T=T_c$ showing deconfinement. At $T=0$ the string tension increases linearly with $m_q$ representing strong confinement \cite{Issifu}.  These graphs are the same as the thermal fluctuating glueball mass $m^{*2}_\phi(T)$ in the IR regime because the string tension is precisely the same as the glueball mass in this regime.
   }
\end{figure}

\begin{figure}[H]
  \centering
  \subfloat[Left Panel]{\includegraphics[scale=0.6]{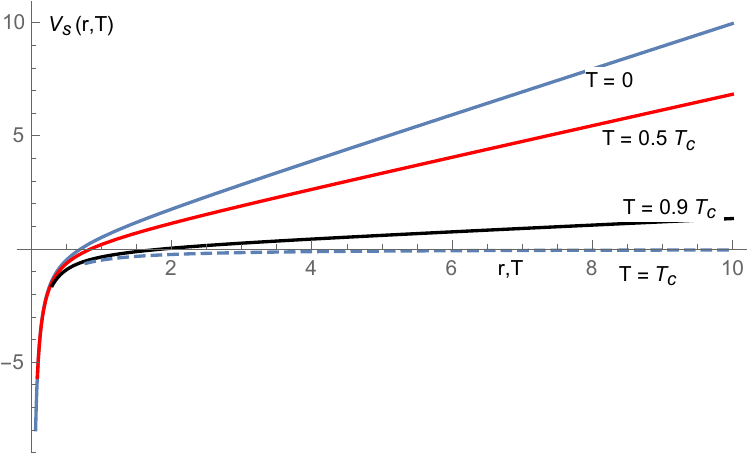}}
  \qquad
  \subfloat[Right Panel]{\includegraphics[scale=0.6]{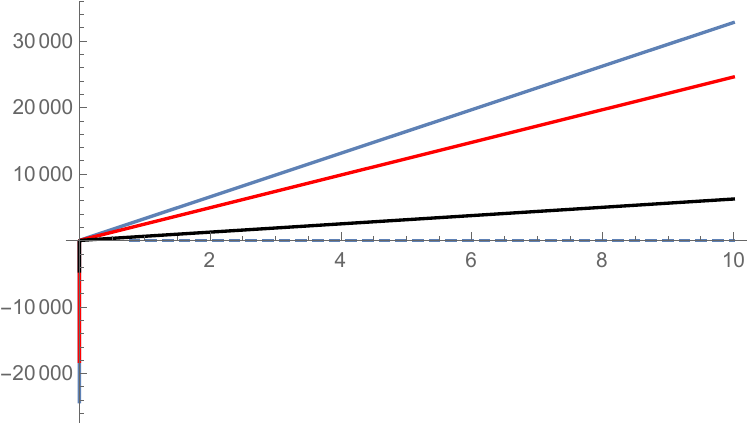}}
  
   \caption{A graph of net potential in the UV regime, $V_{s}(r,T)$, against ($r,T$) for  $m_q=2$ (left) and $m_q\rightarrow\infty$ (right).}
   \label{cn}
   \floatfoot{From the left panel the gradient of the curves decreases with increasing $T$ from $T=0$ (blue), $T=0.5T_c$ (red) to $T=0.9T_c$ (black), indicating a decrease in binding of the quarks as the energy of the system is increasing. At $T=T_c$  (dashed) we have a deconfinement phase and restoration of the chiral symmetry \cite{Gupta,Gottlieb}. %The right panel shows an increase in the strength of confinement as $m_q\rightarrow \infty$, for the same temperature variations as in the left panel. %Here, hadronization occurs due to an excessive increase in mass but the 'break away' quark masses remain confined to each other and to the massive source quark $m_q$, for $T=0$ to $T=0.9T_c$.
   }
\end{figure}

\begin{figure}[H]
  \centering
  \subfloat[Left Panel]{\includegraphics[scale=0.6]{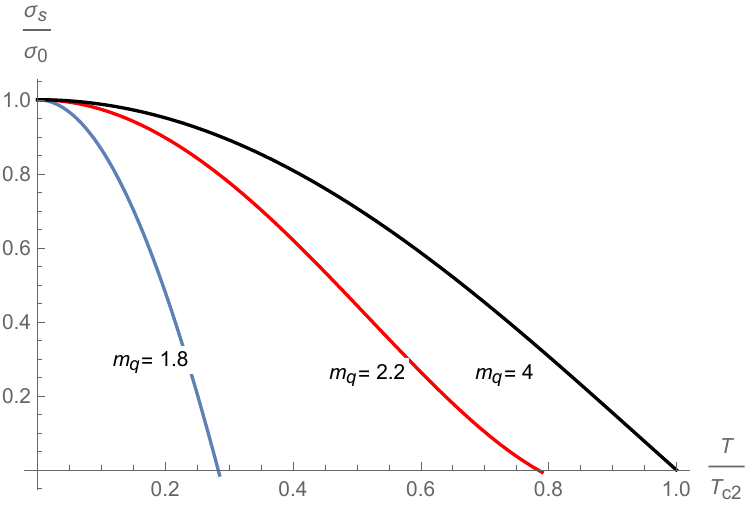}}
  \qquad
  \subfloat[Right Panel]{\includegraphics[scale=0.6]{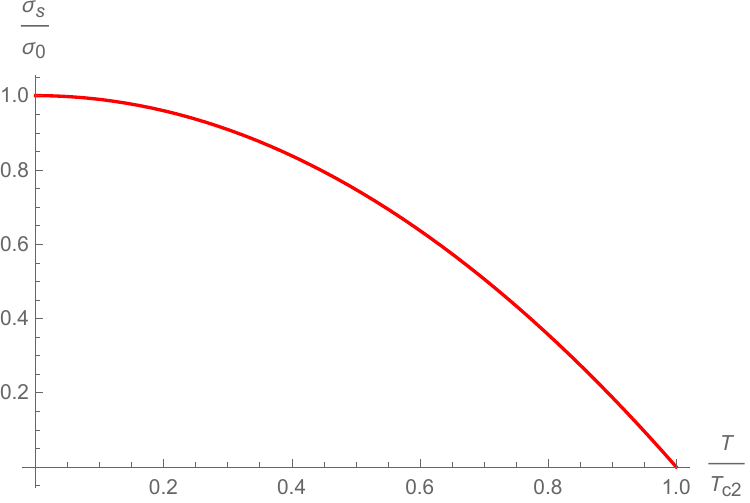}}
  \caption{A graph of string tension, $\sigma_s(T)/\sigma_0$, in the UV regime against $T/T_{c2}$ for different values of $m_q$ (left) and $m_q\rightarrow\infty$ (right).}
   \label{st1}
\floatfoot{This graph follows the same behavior as discussed in Fig.~\ref{st}. However, in this regime the string breaking is explicit as the curves intercept the $T/T_{c2}$ axis at different points. That notwithstanding, the bond states of the light and the heavy quarks will dissolve at $T=T_{c2}$. %but the only difference is that the magnitude of the string tension here is greater than that in Fig.~\ref{st} due to the magnitude of $m_q$. This is particularly needed in this regime to keep the particles in a confined state due to its high energy nature. 
Also, the fluctuating glueball mass $m^2_s(T)$ in this regime is related to the string tension by $m_s(T)=\sqrt{3\sigma_s(T)}$.}
\end{figure}

\begin{figure}[H]
  \centerline{\includegraphics[scale=0.7]{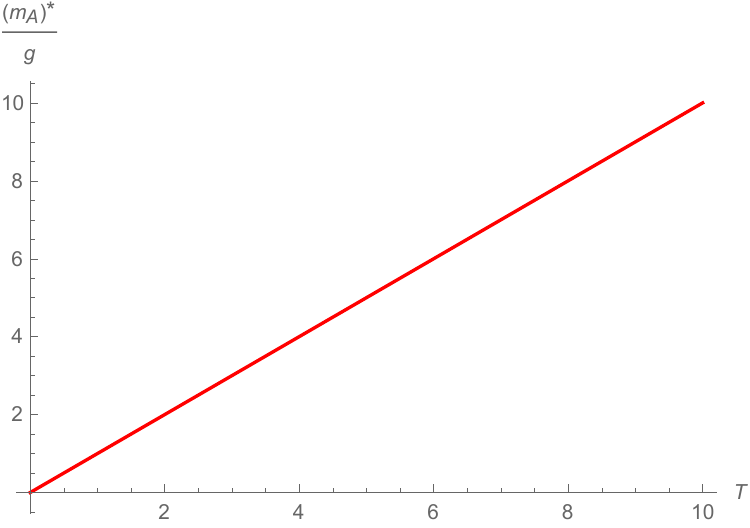}}
  \caption{A graph of fluctuating gluon mass $m^*_A/g$ with temperature $T$.}
   \label{gm}
    \floatfoot{Increase in temperature increases the screening mass.}
\end{figure}

\begin{figure}[H]
  \centerline{\includegraphics[scale=0.7]{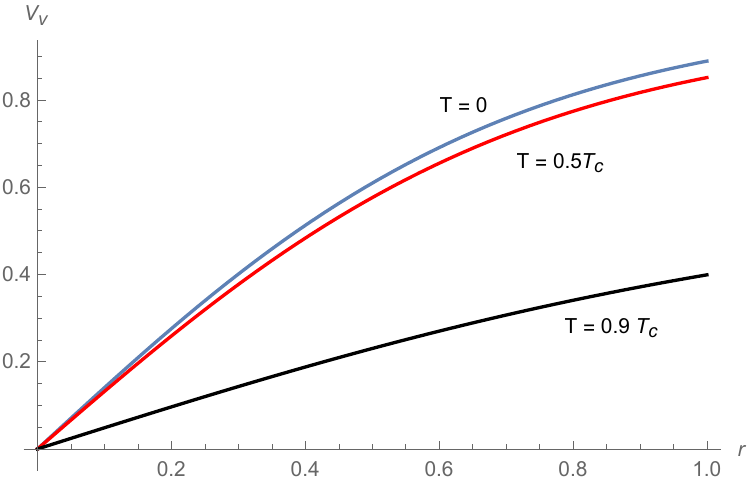}}
  \caption{A graph of vector potential $V_v(T,r)$ against $T,r$ for specific values of $T$.}
   \label{vp}
    \floatfoot{This graph is similar to Fig.~\ref{pt} for $m_q=0$, $T=0$(blue), $T=0.5T_c$(red) and $T=0.9T_c$ (black). This means that the gluons remain confined even if the quark mass is `removed' ($m_q=0$) after confinement. This is known as {\it chromoelectric flux} confinement.}
\end{figure}

\begin{figure}[H]
\centering
  \subfloat[Left Panel]{\includegraphics[scale=0.6]{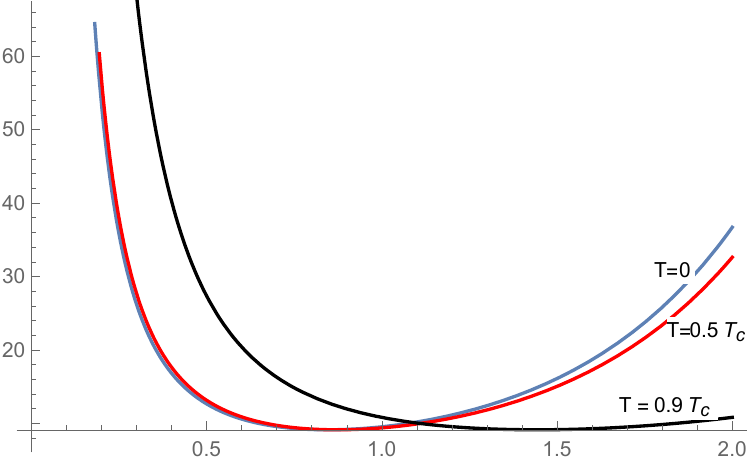}}
   \qquad
   \subfloat[Right Panel]{\includegraphics[scale=0.6]{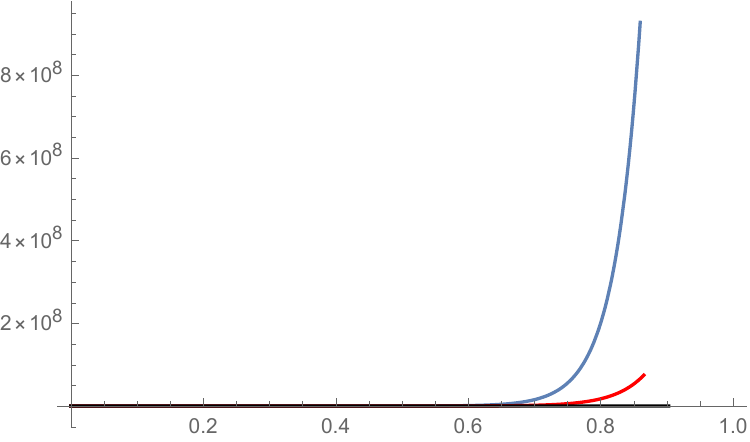}}
  
   \caption{A graph of scalar potential energy, $S(r,T)$, against $r,T$ for $m_q=0.1$(left) and $m_q\rightarrow\infty$(right).}
   \label{sp}
\floatfoot{Confinement is stronger with increasing depth of the curves \cite{Issifu,Bali1}. Thus, the smaller the minima of the curves the more condensed the {glueballs} and stronger the confinement. The minima increases from $T=0$(blue), $T=0.5T_c$(red) to $T=0.9T_c$(black). %The graph in the right panel is for $m_q\rightarrow\infty$. %From both graphs, we find that the curves decreases from left towards the minima and starts rising to the right representing screening of sea quarks \cite{Issifu,Bali1}. Generally, the particles get more confined at lower temperatures and the vice versa. However, we find that for an infinite mass limit (right panel) as the mass is increased all the curves decreases from left towards the minimum and gets flat indicating hadronization. After the hadronizations the red curve ($T=0.5T_c$)  followed by the black curve ($T=0.9T_c$) rises slowly to the right while the blue curve ($T=0$) rises sharply. Thus, screening decreases with increasing temperature.
}
\end{figure}

\begin{figure}[H]
  \centering
  \subfloat[Left Panel]{\includegraphics[scale=0.6]{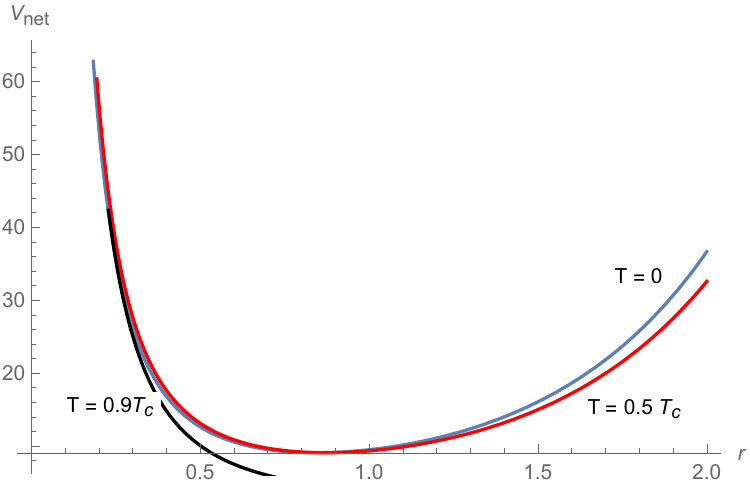}}
  \qquad
  \subfloat[Right Panel]{\includegraphics[scale=0.6]{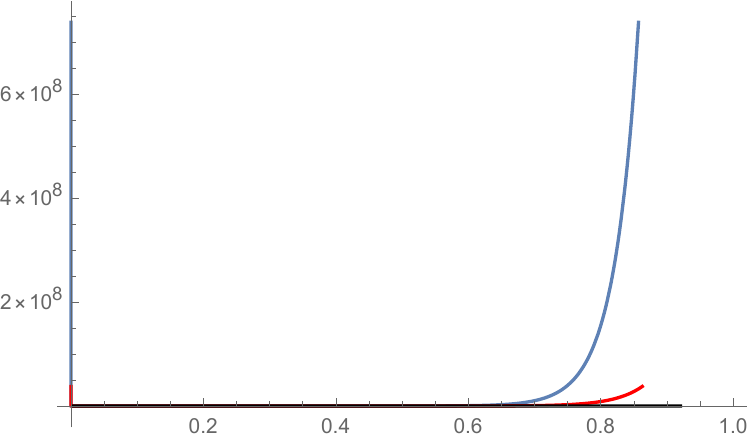}}
  
   \caption{A graph of net potential energy, $V_{\text{net}}(r,T)$, against $r,T$ for $m_q=0.1$ and an infinite $m_q$.}
   \label{np}
\floatfoot{The magnitude of the net potential decreases with increasing temperature, from $T=0$ (blue), $T=0.7T_c$(red) to $T=0.9T_c$(black). %only discrepancy here is the addition of the vector potential energy which strengthen the confinement and minimizes the screening effect.
}
\end{figure}

\begin{figure}[H]
  \centerline{\includegraphics[scale=0.7]{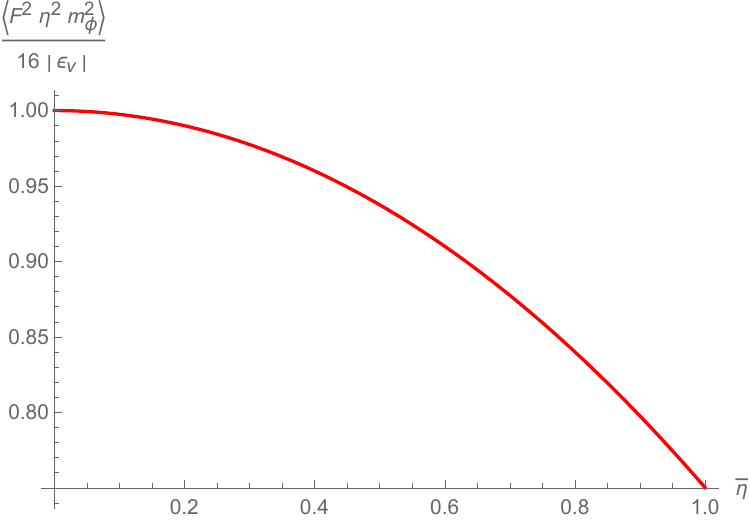}}
  \caption{A graph of gluon condensate $\langle(m^2_\phi\eta^2)F^{\mu\nu}F_{\mu\nu}\rangle/(16|\epsilon_v|)$ against $\bar{\eta}$.}
   \label{gc}
    \floatfoot{The condensation has its maximum value at $\bar{\eta}=0$ and reduces steadily with increasing $\bar{\eta}$ until it vanishes. %Hence, the glueball field $\chi$ has two possible solutions i.e. $\bar{\chi}=0$ and $\bar{\chi}=2/\phi_0$.
    }
\end{figure}

\begin{figure}[H]
  \centerline{\includegraphics[scale=0.7]{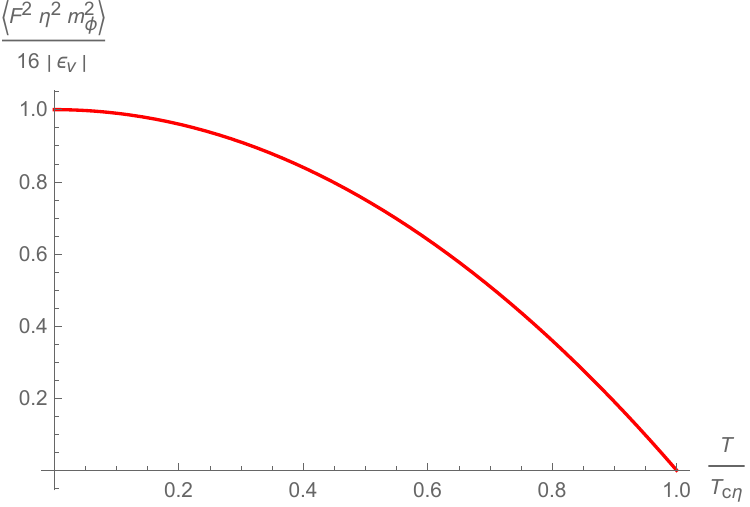}}
  \caption{A graph of gluon condensate $\langle(m^2_\phi\eta^2)F^{\mu\nu}F_{\mu\nu}\rangle/(16|\epsilon_v|)$ against $T/T_{c\eta}$.}
   \label{gc1}
    \floatfoot{The condensate reduces sharply with increasing temperature until it vanishes at $T=T_{c\eta}$.}
\end{figure}

\begin{figure}[H]
  \centerline{\includegraphics[scale=0.7]{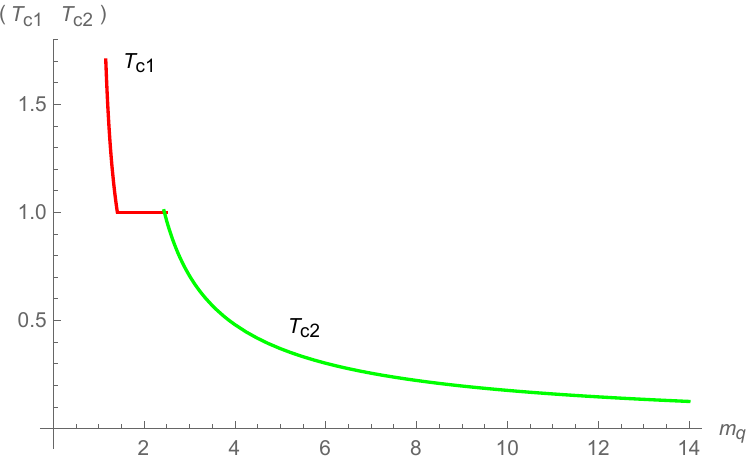}}
  \caption{A phase diagram for $T_{c1}$ and $T_{c2}$ against quark mass $m_q$.}
   \label{srg2}
    \floatfoot{We show a transition between the IR regime corresponding to lighter quark masses to the UV regime corresponding to heavier quark masses. {The non-physical behavior of $T_{c1}$ at masses $m_q\,\leq\,1$ is the regime where confinement of light quarks are observed in the IR regime. The curve becomes constant near $T_{c1}\simeq 1$, the point where the string tension in the IR regime gets saturated and begins to degenerate. Consequently, the IR regime can take masses within $0\,\leq\,m_q\,\leq\,1$, beyond this threshold hadronization sets in and the string tension starts decaying. On the other hand, the UV regime takes in heavier quark masses (green curve), $m_q>\sqrt{3}$. The non-physical behavior observed for $m_q\leq\sqrt{3}$ corresponds to the QGP regime with negative and constantly decaying $\sigma_s$. As $m_q$ increases, $T_{c2}$ decreases and the confinement becomes stronger.}}
\end{figure}

\begin{figure}[H]
  \centering
  \subfloat[Left Panel]{\includegraphics[scale=0.6]{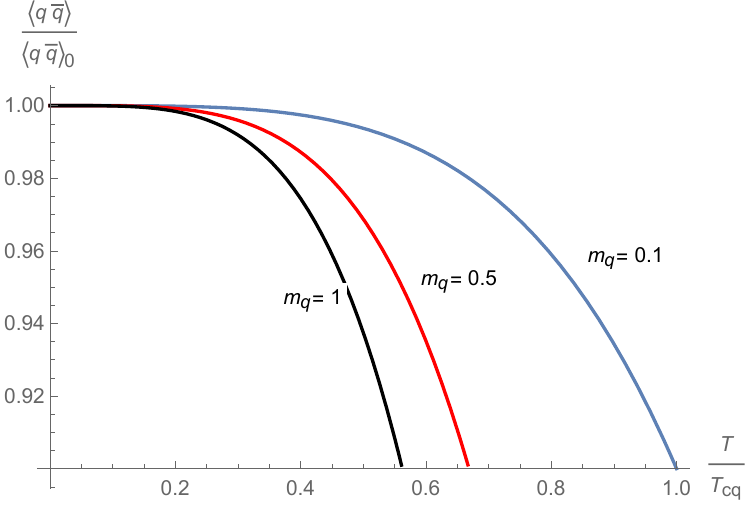}}
  \qquad
  \subfloat[Right Panel]{\includegraphics[scale=0.6]{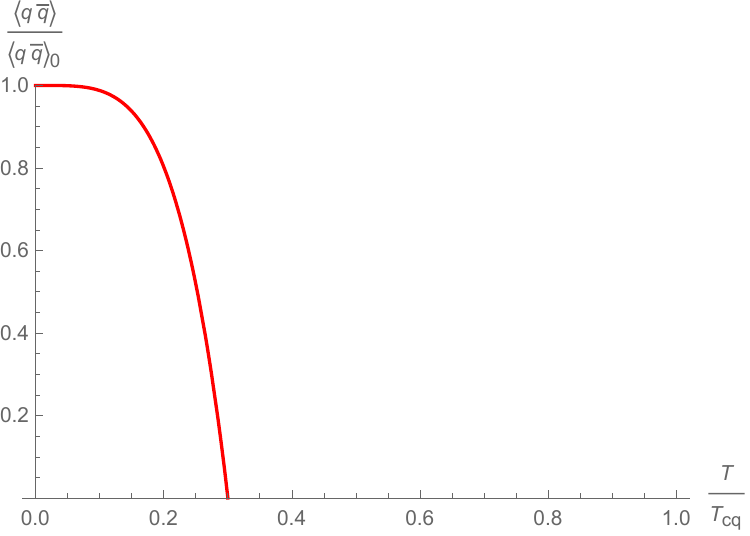}}
  \caption{A graph of chiral condensate against $T/T_{cq}$ for different values of $m_q$ (left) and $m_q\rightarrow\infty$ (right).}
   \label{ch1}
   \floatfoot{An increase in quark mass courses the condensate to reduce sharply and eventually vanish at $T/T_{cq}$ \cite{Jankowski,Borsanyi}. This behavior is opposite that of the string tension in Figs.~\ref{st} and \ref{st1} as expected in QCD lattice simulations \cite{Schmidt}. Consequently, confinement increases with increasing quark mass while the chiral condensate decreases with increasing quark mass.}
\end{figure}

Some of the major results obtained are displayed on a graph to make it easy to see their behavior. Since we have extensively studied and discussed the corresponding results for $T=0$ in \cite{Issifu}, we will concentrate on the results with temperature fluctuations. We plot the {glueball potential} Eq.(\ref{tf8a}) and its behavior with temperature in Fig.~\ref{pt1}. The confining potential Eq.(\ref{tf8}) in IR regime for finite and infinite $m_q$ are plotted in Fig.~\ref{pt} and its string tension Eqs.(\ref{tf9}) plotted in Fig.~\ref{st}. The potential in UV regime Eq.(\ref{tf10}) is plotted in Fig.~\ref{cn} displaying how Cornell-like potential obtained varies with temperature for finite and infinite quark mass limits and their string tension in Eq.(\ref{tf11}) displayed in Fig.~\ref{st1}. The gluon mass Eq.(\ref{gb6b}) which possess all the characteristics of Debye mass is displayed in Fig.~\ref{gm}. The vector potential Eq.(\ref{v5}) which represents {\it chromoelectric flux} confinement is displayed in Fig.~\ref{vp}. The scalar potential energy Eq.(\ref{tf13}) and the corresponding net potential energy Eq.(\ref{s1}) for finite and infinite $m_q$ are plotted in Figs.~\ref{sp} and \ref{np} respectively. The gluon condensates calculated in Eqs.(\ref{vc2}) and (\ref{vc4}) are also displayed in Figs.~\ref{gc} and \ref{gc1} respectively. The color dielectric function in Eq.(\ref{a10}) represents the {glueball potential}. We have higher {glueball condensation} when $G(r)\rightarrow 0$, so it follows the same discussions as Figs.~\ref{sp} and \ref{np}. Also, increase in quark mass $m_q$ will lead to increase in {glueball condensation} \cite{Sen1}. We explored a phase transition from the low energy IR regime to the high energy UV regime by studying the characteristics of $T_{c1}$ and $T_{c2}$ in Eqs.(\ref{phs1}) and (\ref{phs3a}) displayed in Fig.~\ref{srg2}. We find that the critical temperatures decrease with an increase in quark mass thereby increasing the strength of confinement. We observed that the light quarks that are confined in the IR regime $0\,\leq\, m_q\,\leq 1$ are relegated to the QGP phase in the UV regime which confines quarks with $m_q\,>\,\sqrt{3}$. Finally, the chiral condensate $\langle\bar{q}q\rangle$ was calculated in Eq.(\ref{cc5b}) and displayed in Fig.~\ref{ch1}.


%\begin{figure}[H]
 % \centering
 % \subfloat[Left Panel]{\includegraphics[scale=0.5]{string5}}
 % \qquad
 % \subfloat[Right Panel]{\includegraphics[scale=0.5]{string5a}}
 % \caption{Normalized string tension, $\sigma_c/\sigma_0$, against $T/T_c$ for different values of $m_q$ (left) and $m_q\rightarrow\infty$ (right).}
  % \label{string}
  % \floatfoot{The temperature at which we expect the bound states of quarks to dissolve into a 'soup' of their constituents is $T=T_c$, as set by the potential models developed above. Thus, the lower the temperature the longer the life time of the corresponding bound states. As a results, heavier quarks correspond to lower temperatures therefore, heavy quarks will have longer life time in the presence of temperature than the lighter ones \cite{Digal,Karsch,Antonov}.}
%\end{figure}

%\begin{figure}[H]
 % \centering
  %\subfloat[Left Panel]{\includegraphics[scale=0.5]{string6}}
  %\qquad
 % \subfloat[Right Panel]{\includegraphics[scale=0.5]{string6a}}
 % \caption{Normalized string tension, $\sigma_s/\sigma_0$, against $T/T_c$ for different values of $m_q$ (left) and an infinite $m_q$ (right)}
   %\label{string1}
  % \floatfoot{This graph follows similar discussion as Fig.~\ref{string}, lighter bound states are more likely to dissolve faster than the heavier ones.}
%\end{figure}

%\begin{figure}[H]
  %\centerline{\includegraphics[scale=0.7]{stringf2}}
  %\caption{A graph of $\sigma_s/\sigma_0$ against $T/T_{c2}$.}
   %\label{srg1}
    %\floatfoot{In this regime, we find that the string tension breaks quickly for lighter quarks than heavier ones with temperature as expected.}
%\end{figure}

%\begin{figure}[H]
 % \centerline{\includegraphics[scale=0.7]{chiral}}
 % \caption{A graph of chiral condensate against $T/T_{cq}$.}
  % \label{ch}
   % \floatfoot{The condensate reduces with increasing temperature indicating reduction in the strength of confinement until it vanishes at $T=T_{cq}$ indicating restoration of chiral symmetry \cite{Jankowski,Borsanyi}.}
   %\end{figure}

%\begin{figure}[H]
 % \centerline{\includegraphics[scale=0.7]{chiral1}}
 % \caption{A graph of chiral condensate against $T/T_c$ for different values of $m_q$.}
  % \label{gm1}
   % \floatfoot{The condensate decreases with increasing quark mass.}
%\end{figure}

\acknowledgments
We would like to thank CNPq, CAPES and CNPq/PRONEX/FAPESQ-PB (Grant no. 165/2018),  for partial financial support. FAB also acknowledges support from CNPq (Grant no.  312104/2018-9).

\end{document}